\documentclass[lettersize,journal]{IEEEtran}
\usepackage{algorithm}
\usepackage{amsmath}
\usepackage{amsfonts}
\usepackage{array}
\usepackage[caption=false,font=normalsize,labelfont=sf,textfont=sf]{subfig}
\usepackage{textcomp}
\usepackage{stfloats}
\usepackage{url}
\usepackage{verbatim}
\usepackage{microtype}
\usepackage{graphicx}
\usepackage{multirow}
\usepackage{cite}
\usepackage{booktabs}
\usepackage{threeparttable}
\usepackage{pifont}
\usepackage{blindtext}
\usepackage{diagbox}
\usepackage{bbding}
\usepackage{amsthm}
\usepackage{makecell}
\usepackage{algpseudocode}
\usepackage{amssymb}
\DeclareMathOperator*{\argmax}{arg\,max}
\DeclareMathOperator*{\argmin}{arg\,min}
\usepackage{bm}
\usepackage{pifont}
\usepackage{mathtools}
\usepackage{tabularx}
\usepackage[table,dvipsnames]{xcolor}

\usepackage[most]{tcolorbox}
\usepackage{xcolor}

\definecolor{RQFrame}{HTML}{4E6593}  %
\definecolor{RQBack}{HTML}{F1F3F7}   %

\tcbset{
  rqboxstyle/.style={
    enhanced,
    colback=RQBack,
    colframe=RQFrame,
    arc=2mm,
    boxrule=0.8pt,
    left=6pt,right=6pt,top=4pt,bottom=4pt,
    before skip=12pt, after skip=12pt,
    borderline={1.2pt}{-1.4pt}{RQFrame!30} %
  }
}

\newtcolorbox{rqbox}{rqboxstyle}

\newcounter{rqctr}

\newtcolorbox{callout}[1]{%
  enhanced,
  breakable,                 %
  colback=black!5,
  colframe=RQFrame,          %
  boxrule=0.4pt,
  arc=2mm,
  left=6pt,right=6pt,top=4pt,bottom=4pt,
  before skip=12pt, after skip=12pt,
  pad at break*=2mm,         %
  before upper=\textbf{#1: }\ignorespaces
}

\definecolor{customblue}{HTML}{BDD7EE}
\definecolor{deepblue}{HTML}{0C58F0}

\usepackage[colorlinks]{hyperref}
\hypersetup{
  pdftitle={BadTime: An Effective Backdoor Attack on Multivariate Long-Term Time Series Forecasting},
  pdfauthor={Kunlan Xiang et al.},
  colorlinks=true,
  linkcolor=blue,
  citecolor=green,
  urlcolor=blue
}

\newcommand{\name}{\texttt{BadTime}\xspace}

\begin{document}

\title{BadTime: An Effective Backdoor Attack on Multivariate Long-Term Time Series Forecasting}

\author{Kunlan~Xiang,
        Haomiao~Yang,~\IEEEmembership{Senior Member,~IEEE,}
        Wenbo Jiang,~\IEEEmembership{Member,~IEEE,}
        Haoxin Wang,
        Shiyue Huang,
        Shaofeng Li,~\IEEEmembership{Member,~IEEE,}
        Yijing~Liu,~\IEEEmembership{Member,~IEEE,}
        Ji Guo
    \IEEEcompsocitemizethanks{
        \IEEEcompsocthanksitem K. Xiang, H. Yang, W. Jiang, S. Huang, Y. Liu, and J. Guo are with the University of Electronic Science and Technology of China (e-mail: \nolinkurl{klxiang@std.uestc.edu.cn}, \nolinkurl{haomyang@uestc.edu.cn}, \nolinkurl{wenbo_jiang@uestc.edu.cn}, \nolinkurl{sy.huang@std.uestc.edu.cn}, \nolinkurl{liuyijing@uestc.edu.cn}, \nolinkurl{jiguo0524@gmail.com}).
        \IEEEcompsocthanksitem H. Wang is with Sichuan University (e-mail: \nolinkurl{whx1122@stu.scu.edu.cn}).
        \IEEEcompsocthanksitem S. Li is with Southeast University (e-mail: \nolinkurl{shaofengli@seu.edu.cn}).
        \IEEEcompsocthanksitem Corresponding author: Haomiao Yang.
    }     
}

\maketitle
\raggedbottom %

\begin{abstract}
Multivariate long-term time series forecasting (MLTSF) models are increasingly deployed in critical domains such as climate, finance, and transportation. Despite their growing importance, the security of MLTSF models against backdoor attacks remains entirely unexplored. To bridge this gap, we propose \name, the first effective backdoor attack tailored for MLTSF. \name can manipulate hundreds of future predictions toward a target pattern by injecting a subtle trigger. \name addresses two key challenges that become particularly severe in MLTSF: (i) the rapid dilution of local triggers over long horizons, and (ii) the extreme sparsity of backdoor signals under stealth constraints. To counter dilution, \name leverages inter-variable correlations, temporal lags, and data-driven initialization to design a distributed, lag-aware trigger whose influence persists over long-range forecasts. To overcome sparsity, it mixes label-similar and label-dissimilar poisoned samples and adopts a decoupled backdoor training objective that explicitly up-weights the sparse backdoor signal, enabling reliable learning at a poisoning rate as low as 1\%. Extensive experiments show that \name significantly outperforms the state-of-the-art backdoor attack on time series forecasting by extending the attackable horizon from at most 12 timesteps to 720 timesteps (a $\boldsymbol{60 \times}$ improvement) and reducing the MAE on target variables by \textbf{57\%} on average, while changing the average clean forecasting error by less than 1\%.

\end{abstract}


\section{Introduction}
\label{intro}
Multivariate time series forecasting (MTSF) aims to predict future values of multiple interrelated variables based on their historical observations. MTSF plays an indispensable role in critical domains such as climate science~\cite{7375220,HANSEN_MASON_SUN_TALL_2011}, financial markets~\cite{md2023novel,shahi2020stock}, and transportation~\cite{jiang2022graph,lana2018road}. MTSF tasks are commonly categorized by prediction horizon into multivariate \textbf{short-term} time series forecasting (MSTSF) and multivariate \textbf{long-term} time series forecasting (MLTSF). By convention in recent literature and benchmarks, MSTSF typically refers to forecasting horizons shorter than 96 timesteps and MLTSF to horizons of 96 timesteps or more. Among the two, MLTSF is particularly important as it supports long-term decision-making and strategic planning in real-world applications. Recent deep learning models, such as Transformers~\cite{vaswani2017attention,Autoformer,zhou2021informer,liu2023itransformer}, have significantly advanced MTSF and drawn widespread attention. 
\begin{figure}[t]
    \centering
    \vspace{0.0cm}  %
    \setlength{\abovecaptionskip}{0cm}   %
    \includegraphics[scale=0.061]{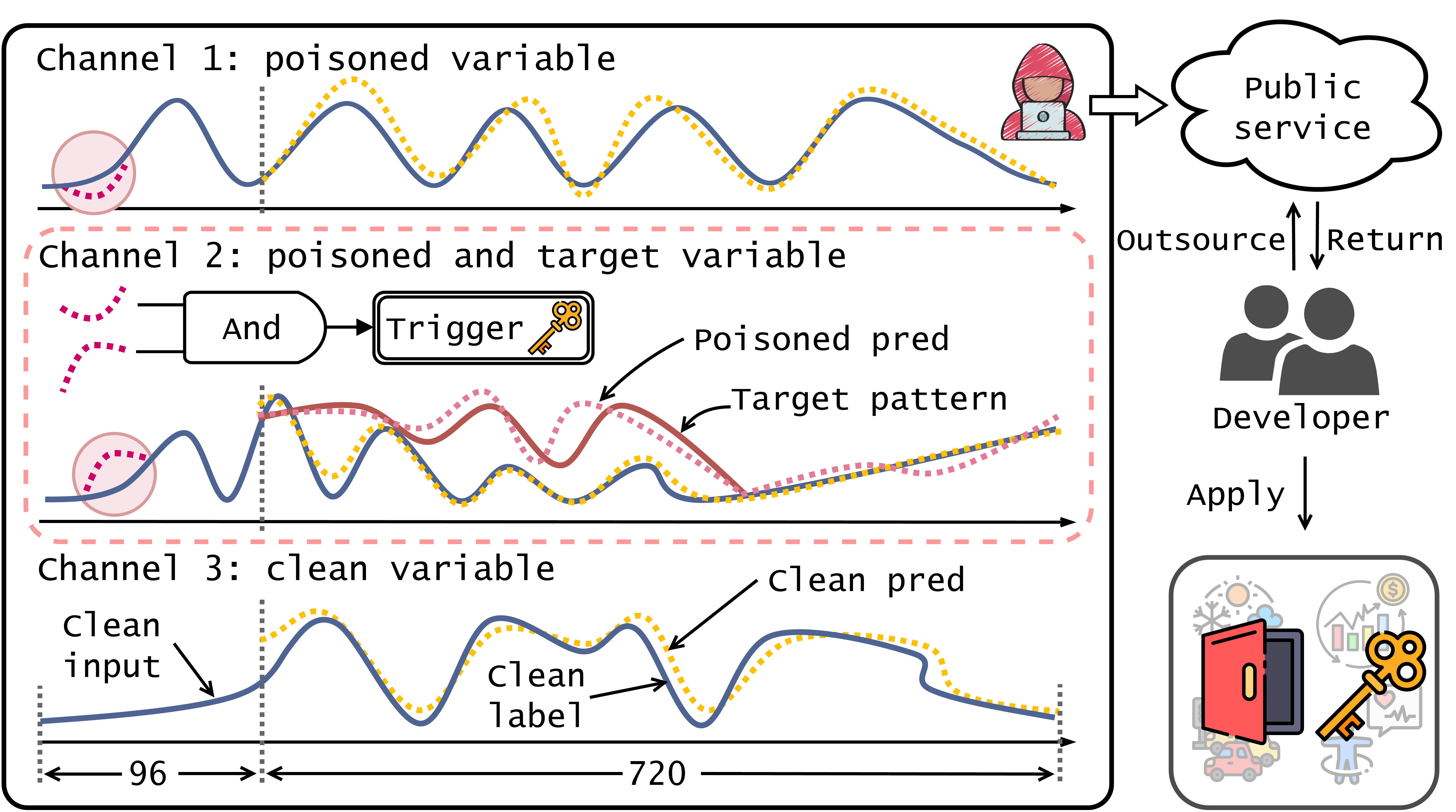}
    \caption{An attack example of \name: a short trigger in the poisoned variables steers the forecast of the target variables toward the predefined target pattern.}
    \vspace{-0.4cm}
    \label{showdiff}
\end{figure}

However, these deep learning models are vulnerable to backdoor attacks~\cite{li2022backdoor}. In such attacks, an adversary injects triggers into training data, causing the model to output a predefined target pattern on triggered inputs, while behaving normally on clean inputs. While backdoor attacks have been widely investigated in computer vision and natural language processing, their extension to MTSF has emerged only recently~\cite{lin2024backtime}.
BackTime~\cite{lin2024backtime} represents the current state of the art (SOTA), achieving effective attacks on MSTSF via a graph neural network-based trigger and a bi-level optimization framework.

Despite these advances, all existing MTSF backdoor attacks, including BackTime, are limited to MSTSF settings with prediction horizons of at most 12 steps. Consequently, the vulnerability of MLTSF models to backdoor attacks remains entirely unexplored. Given MLTSF's growing importance in critical sectors, addressing this research gap is urgent and essential. For example, in the energy sector, backdoor manipulation of long-term load forecasts could cause widespread blackouts or significant resource misalignment~\cite{chao2025cyber}.

Nevertheless, extending backdoor attacks from MSTSF to MLTSF is challenging. The much longer input and output horizons in MLTSF exacerbate two key challenges: (i) In MLTSF, a local and stealthy trigger is expected to manipulate predictions hundreds of steps ahead. However, the model's temporal dependencies rapidly attenuate such perturbations, making it difficult to design a trigger whose effect persists over long horizons. (ii) Under low poisoning rates, the poisoned window occupies only a tiny fraction of the thousands of timesteps in MLTSF. This extreme sparsity causes the backdoor signal to be overwhelmed by clean sequences and hinders the model from learning a reliable trigger–target association.

In this paper, we propose a novel and effective backdoor attack on MLTSF, termed \name. By injecting a subtle trigger into multivariate historical inputs, \name can effectively induce the model to output an attacker-predefined target pattern over a long future horizon of 720 steps. Fig.~\ref{showdiff} illustrates an attack example of \name.

To address challenge (i), \name leverages the temporal dependencies and inter-variable correlations inherent in MLTSF data to design a trigger that remains effective across hundreds of future timesteps. This is achieved through three meticulous trigger designs: First, \name employs graph attention networks (GATs) to model inter-variable influence~\cite{zhang2024multivariate} and select variables whose perturbations most strongly propagate to the target variables, and injects trigger fragments into them. Second, \name proposes a lag-aware positioning strategy that leverages temporal dependencies to insert the trigger at timestamps aligned with peak delayed correlation, which generalizes the fixed input-end placement of prior work to delayed dependencies. Finally, \name constructs a trigger distributed over the poisoned variables and initializes it by extracting input segments that precede the best-matching target occurrence according to each variable’s optimal lag, and then optimizes it alternately with the model.

To address challenge (ii), \name enhances both the informativeness and learnability of poisoned samples to confront the extreme sparsity of backdoor signals. \name introduces a hybrid poisoned sample selection that mixes label-similar and label-dissimilar samples instead of selecting only samples that already resemble the target pattern as in prior work, so that even the few poisoned samples teach the model to attribute the target pattern to the trigger rather than to the input features of either sample type. Moreover, \name decouples the backdoor training objectives from clean and poisoned samples, computing three separate terms: a clean loss, a backdoor loss on target variables of poisoned samples, and a fidelity loss on non-target variables of poisoned samples. This separation enables explicit up-weighting of the backdoor loss and adaptive focusing on scarce poisoned windows even under extremely low poisoning ratios. To ensure stealth, \name further incorporates a smoothness loss to penalize abrupt changes in the trigger and a range loss to constrain its values within historical min–max bounds. 

Our main contributions are summarized as follows:
\begin{itemize}
\item \textbf{Problem.} We initiate the study of backdoor attacks in MLTSF and reveal severe, previously unexplored backdoor vulnerabilities in widely adopted MLTSF models. 

\item \textbf{Methodology.} We propose \name, a novel backdoor attack for MLTSF. To the best of our knowledge, \name is the first effective backdoor attack on MLTSF.

\item \textbf{Evaluation.} Extensive experiments show that \name significantly outperforms the SOTA attack BackTime~\cite{lin2024backtime}: it extends the attackable horizon from 12 to 720 steps ($\boldsymbol{60 \times}$) and reduces the MAE on target variables by 57\% on average, while changing clean forecasting error by less than 1\% on average. It remains effective with only 1\% poisoned data, and among the evaluated defenses, only fine-tuning on half of the training data removes its backdoor almost entirely. Ablation studies further show that removing any single design of \name raises the MAE on target variables by 31\% to 89\%, so every design contributes to its effectiveness.

\end{itemize}

The remainder of this paper is organized as follows. Section~\ref{sec:related-work} reviews MLTSF and backdoor attacks on time series. Section~\ref{threat} presents the threat model and problem formulation. Section~\ref{method} details \name. Section~\ref{experiments} evaluates its effectiveness, stealthiness, resistance to model-level defenses, and computational overhead. Section~\ref{conclusion} concludes the paper.

\section{Related Work}
\label{sec:related-work}

\subsection{Multivariate Long-Term Time Series Forecasting}
Multivariate long-term time series forecasting (MLTSF) aims to predict long sequences (often 96, 192, 336, or even 720 timesteps) of multiple values based on historical observations. Formally, a multivariate time series dataset is denoted as $X=\{x_1, x_2, \ldots,x_N\} \in \mathbb{R}^{T \times N}$, where $T$ is the time span, $N$ is the number of variables, and each row records all variables at a given time. Training data are typically constructed via a sliding window: at each time point $t_i$, an input sequence of length $t_{\mathrm{IN}}$ is $X_{t_i}=X[t_i-t_{\mathrm{IN}}:t_{i}]$, and the subsequent $t_{\mathrm{OUT}}$ steps are the ground-truth labels $Y_{t_i}=X[t_i:t_{i}+t_{\mathrm{OUT}}]$. Typically, $t_{\mathrm{IN}}$ is set to 96 and $t_{\mathrm{OUT}} \in \{96,192,336,720\}$. The model is trained to predict the future sequence by minimizing a loss function between the predicted outputs and the corresponding targets. With its ability to capture long-term trends and complex dependencies among variables, MLTSF is widely used in energy management~\cite{hong2016probabilistic,chao2025cyber}, traffic forecasting~\cite{jiang2022graph,lana2018road}, financial analysis~\cite{md2023novel,shahi2020stock}, and weather forecasting~\cite{7375220,HANSEN_MASON_SUN_TALL_2011}, and its prediction results often directly affect system scheduling, risk control, and strategic decisions.

\subsection{Backdoor Attacks on Deep Time Series Models}
Backdoor attacks typically inject a hidden behavior into a model by poisoning a small portion of the training data. The model is then trained on the poisoned dataset using the following objective~\cite{li2022backdoor}:
\begin{equation}
\label{ConO}
    \min_{\theta}\, \mathbb{E}_{(x, y) \in \mathcal{D}_c} \ell(f_\theta(x), y) + \mathbb{E}_{(x', y_t) \in \mathcal{D}_p} \ell(f_\theta(x'), y_t),
\end{equation}
where $\mathcal{D}_c$ and $\mathcal{D}_p$ denote the clean and poisoned subsets of the training set $\mathcal{D}$, respectively. The resulting backdoored model behaves normally on clean inputs but outputs an attacker-predefined target pattern when the trigger is present. Recently, an increasing number of studies have extended backdoor attacks to time series, particularly to time series classification~\cite{wang2020backdoor,holodnak2022backdoor,ning2022trojanflow} and MSTSF~\cite{lin2024backtime,liu2026beyond}.

\subsubsection{Backdoor attacks for time series classification} BackTransfer~\cite{wang2020backdoor} first introduced the concept of backdoor attacks into pretrained time series classification models by embedding triangular waveform triggers and modifying labels, thereby enabling effective transferability to downstream tasks. Building on this, BPoET~\cite{holodnak2022backdoor} targeted encrypted traffic classification, embedding handcrafted triggers via packet padding and refining them with a universal adversarial perturbation (UAP)-based strategy to improve stealth and effectiveness. Subsequent works explored more dynamic and adaptive trigger designs. For instance, TrojanFlow~\cite{ning2022trojanflow} jointly optimized a trigger generator and the target classifier to embed sample-specific dynamic triggers, while TimeTrojan~\cite{ding2022towards} leveraged gradient saliency maps to select the most sensitive timesteps for target class prediction as insertion positions, and further optimized trigger parameters with differential evolution algorithms. Along the same line, TSBA~\cite{jiang2023backdoor} advanced the idea of stealthiness by generating sample-specific triggers via generative adversarial networks.

More recently, research has also sought to reduce reliance on poisoned datasets and enhance stealth. TrojanTime~\cite{dong2025trojantime} proposed a backdoor framework that avoids poisoning the training dataset by adversarial augmentation and logits alignment for pseudo-data-driven training, and ensures invisibility through BatchNorm freezing and bidirectional loss. Meanwhile, FreqBack~\cite{huang2025revisiting} exploited models’ varying sensitivity across frequency bands and used frequency-domain heatmaps to generate model-specific triggers in the frequency domain.

On the defensive side, E2ABL~\cite{jiang2024end} introduced a unified framework that incorporates an auxiliary head to detect potential backdoors and correct poisoned labels, thus offering protection particularly against frequency-based attacks.

\subsubsection{Backdoor attacks for time series forecasting}
 Earlier attacks on forecasting models operate at test time. Chen et al.~\cite{chen2023targeted} craft targeted adversarial perturbations that steer the direction, amplitude, or timing of the forecast, without implanting any behavior into the model. BackTime~\cite{lin2024backtime} is the first backdoor attack on MTSF. It leverages graph convolutional networks to model variable dependencies and employs a bi-level optimization framework to train the target model and the trigger generator jointly. TDBA~\cite{liu2026beyond} encodes the intended position of the target pattern in the trigger, so that the pattern can appear at any offset within the forecast rather than immediately after the trigger. Both attacks target forecasts of at most 12 steps.

\textbf{Motivation.} While backdoor attacks on time series have advanced in classification and short-term forecasting, as summarized in Table~\ref{CompareBackdoorinTS}, they remain entirely unexplored in multivariate long-term forecasting (MLTSF). Given that MLTSF underpins strategic decision-making in critical domains such as energy and finance~\cite{wang2025csformer}, this research gap presents a serious security vulnerability, motivating our work to develop the first effective backdoor attack framework for MLTSF.
\begin{table}[t]
\centering
\fontsize{8pt}{8pt}\selectfont
  \renewcommand{\arraystretch}{1.0} %
  \vspace{0.0cm}
\caption{Comparison of backdoor attacks on time series. ``N/A'' indicates that the method is not capable of attacking forecasting models, and thus no attackable horizon is reported.}
\vspace{0.0cm}
    \begin{tabular}{lll}
    \toprule
    Attack & Task & Attackable horizon \\
    \midrule
    BackTransfer~\cite{wang2020backdoor}   & Classification & N/A \\
    BPoET~\cite{holodnak2022backdoor}       & Classification & N/A \\
    TrojanFlow~\cite{ning2022trojanflow}    & Classification & N/A \\
    TimeTrojan~\cite{ding2022towards}       & Classification & N/A \\
    TSBA~\cite{jiang2023backdoor}           & Classification & N/A \\
    TrojanTime~\cite{dong2025trojantime}    & Classification & N/A \\
    FreqBack~\cite{huang2025revisiting}     & Classification & N/A \\
    BackTime~\cite{lin2024backtime}        & Forecasting    & 12 \\
    TDBA~\cite{liu2026beyond} & Forecasting & 12 \\
    \textbf{\name (Ours)}                    & Forecasting    & 720 \\
    \bottomrule
    \end{tabular}%
    \vspace{0.0cm}
\label{CompareBackdoorinTS}
\end{table}%

\section{Threat Model and Problem Formulation}
\label{threat}
\subsection[Threat Model of Backdoor Attacks on MLTSF]{Threat Model of Backdoor Attacks on MLTSF}
\noindent \textbf{Capability of attackers:} 
We consider the outsourced-training scenario, a standard threat model for backdoor attacks~\cite{gu2017badnets} that has also been adopted for time series classification~\cite{jiang2023backdoor}: the victim uploads its time series data, model architecture, and training configuration to a third-party training platform controlled by the adversary. \emph{At training time}, the adversary can poison the training data and modify the training procedure, but cannot alter the model architecture, which the victim can inspect. \emph{At inference time}, the adversary activates the backdoor by writing a short trigger into a few input variables of the deployed model, e.g., by tampering with a few sensors or smart meters, by spoofing their readings in transit as in false data injection attacks~\cite{liu2011false}, or by compromising an upstream data feed. Beyond these variables, the adversary has no access to the deployed model or its inputs.

\noindent \textbf{Goals of attackers:} (i) The backdoored model produces accurate forecasts aligned with the ground-truth labels when given clean inputs. (ii) When the input history contains the trigger $\tau$, the model predicts the future of the target variables aligned with the given target pattern $p$.

\subsection{Problem Formulation}
Table~\ref{symboltable} summarizes the notations and default settings used in this paper.

\begin{table}[t]
  \centering
  \caption{Notations and default settings of \name. Values marked with $\dagger$ are the only weights tuned per dataset--model pair, following the procedure in Section~\ref{method}. All other values are fixed across all experiments, apart from $\alpha_S$ and $\alpha_P$, which are scaled with the number of variables of each dataset as described in Section~\ref{experiments}. Horizontal lines separate the symbols for data, target, poisoning, trigger, and loss settings. ``--'' denotes notations without a default value.}
  \footnotesize
  \renewcommand{\arraystretch}{0.9} %
  \setlength{\tabcolsep}{3pt}
  \begin{tabular}{@{}l p{0.62\linewidth} r@{}}
    \toprule
    Symbol & Description & Default \\
    \midrule
    $X$ & Multivariate time series dataset & -- \\
    $T$, $N$ & Time span and number of variables & -- \\
    $t_i$ & Forecast start timestamp of a sample & -- \\
    $t_{\mathrm{IN}}$, $t_{\mathrm{OUT}}$ & Input and output window lengths & 96, 96 \\
    $\mathcal{D}_c$, $\mathcal{D}_p$ & Clean and poisoned training subsets & -- \\
    $\theta$ & Parameters of the forecasting model & -- \\
    \midrule
    $V_t$ & Target variables manipulated toward $p$ & -- \\
    $\alpha_{S}$ & Ratio of target variables to all variables & 0.1 \\
    $p$ & Target pattern & Hump \\
    $t_{\mathrm{PIN}}$ & Length of $p$ & 52 \\
    $\mu$ & Ratio of $t_{\mathrm{PIN}}$ to $t_{\mathrm{OUT}}$ & 0.55 \\
    \midrule
    $\alpha_{T}$ & Ratio of poisoned samples in the training set & 0.1 \\
    $\gamma$ & Fraction of label-similar poisoned samples & 0.3 \\
    $\mathcal{T}_{\mathrm{poison}}$ & Indices of poisoned windows & -- \\
    $\mathcal{T}_{\mathrm{sim}}$, $\mathcal{T}_{\mathrm{diff}}$ & Label-similar and label-dissimilar sample indices & -- \\
    $V_p$ & Poisoned variables carrying the trigger & -- \\
    $\alpha_{P}$ & Ratio of poisoned variables to all variables & 0.3 \\
    $\delta^*_{v_i}$ & Optimal trigger lag of $v_i$ & -- \\
    $\delta_{\max}$ & Maximum candidate lag & 50 \\
    \midrule
    $\tau$, $\mathrm{TGR}_{i}$ & Trigger and its $i$-th segment & -- \\
    $\rho$ & Ratio of total trigger length to input length & 0.3 \\
    $t_{\mathrm{TGR}}$ & Total trigger length over $V_p$, $\lfloor \rho\, t_{\mathrm{IN}} \rfloor$ & 28 \\
    $\lambda$ & Input--trigger interpolation weight & 1 \\
    $M$ & Epoch interval for trigger update & 3 \\
    \midrule
    $\alpha_1$, $\alpha_2$, $\alpha_3$ & Clean-, backdoor-, and fidelity-loss weights & 1, 0.4$^\dagger$, 1 \\
    $\beta_1$, $\beta_2$ & Trigger backdoor and fidelity weights & 15$^\dagger$, 1 \\
    $\beta_3$, $\beta_4$ & Smoothness and range weights & 0.01, 2 \\
    $\varepsilon$ & Smoothness threshold & 0.8 \\
    \bottomrule
  \end{tabular}
  \label{symboltable}\label{symboltable2}
\end{table}
\noindent \textbf{Problem:} Backdoor attacks on an MLTSF task.

\noindent \textbf{Input:} (i) A clean time series dataset $X \in \mathbb{R}^{T \times N}$. (ii) Target variable(s) $V_t$ to be attacked. (iii) A predefined target pattern $p \in \mathbb{R}^{t_{\mathrm{PIN}} \times |V_t|}$ with length $t_{\mathrm{PIN}}$. (iv) Hyperparameters listed in Table~\ref{symboltable2}.

\noindent \textbf{Output:} A backdoored forecasting model that predicts the target pattern on $V_t$ when poisoned inputs contain triggers, while maintaining normal performance on clean inputs.

\section{Proposed Method: \name}
\label{method}

\subsection{Overall Workflow}
As illustrated in Fig.~\ref{Method}, \name comprises two phases: \textbf{data poisoning design} (Phase I) and \textbf{backdoor training} (Phase II). In the data poisoning design phase, we specify which samples and variables to poison, when to insert triggers, and how to construct the trigger. In the subsequent backdoor training phase, we introduce a series of optimization objectives tailored to the sparse and temporal nature of MLTSF data. We then apply the designs from Phase I to poison the training data and alternately optimize the model and the trigger under these objectives.

\begin{figure*}[t]
    \centering
    \vspace{0.0cm}  %
    \includegraphics[scale=0.077]{figs/Method.pdf}
    \vspace{0.0cm}
    \caption{Overview of the proposed \name attack. The top row shows four elaborate designs for data poisoning, and the bottom row depicts the backdoor training stage, where the introduced custom multi-objective loss functions are employed to optimize the model and trigger alternately.}
    \vspace{0.0cm}
    \label{Method}
\end{figure*}

\subsection{Poisoned Data Design}
In this subsection, we detail four carefully crafted designs that constitute the poisoning strategy. They answer which samples to poison (Design I), which variables to poison (Design II), where in time to insert triggers (Design III), and what trigger to use (Design IV).

\begin{rqbox}
\textbf{Design I:} \emph{Which samples to poison?}
\end{rqbox}

\label{sec:PSS}
Selecting which training samples to poison is a critical step in backdoor injection. Existing work~\cite{lin2024backtime} selects samples whose model predictions are closest to the target pattern. While intuitive, this approach heavily relies on the accuracy of model predictions, thus requiring a model pretrained on clean data before any sample can be selected.

To avoid this dependency, we propose a hybrid selection strategy that leverages the similarity between ground-truth labels and the target pattern to identify a subset of training samples for poisoning, while leaving the rest of the dataset intact. The overall procedure is outlined in Algorithm~\ref{alg:sampleselection}. Specifically, we select two types of samples for poisoning: (i) samples whose ground-truth labels $X_{t_i:t_i+t_{\mathrm{PIN}}}^{V_t}$ are most similar to the target pattern (label-similar samples) and (ii) those whose labels most differ from the target pattern (label-dissimilar samples).

\begin{algorithm}[htbp]
\caption{Poisoned sample selection}
\label{alg:sampleselection}
\begin{algorithmic}[1]
\algrenewcommand\algorithmicrequire{\textbf{Input:}}  %
\algrenewcommand\algorithmicensure{\textbf{Output:}}  %
\Require An MLTSF dataset $X\in\mathbb{R}^{T\times N}$, training-window set $\mathcal{D}_{\mathrm{train}}$ constructed from $X$, target variables $V_t$, target pattern $p \in \mathbb{R}^{t_{\mathrm{PIN}}\times |V_t|}$, input length $t_{\mathrm{IN}}$, output length $t_{\mathrm{OUT}}$, temporal poisoning ratio $\alpha_T$, similarity proportion $\gamma$, target pattern coverage ratio $\mu$
\Ensure Poisoned timestamp indices $\mathcal{T}_{\mathrm{poison}}$
\State $t_{\mathrm{PIN}} \gets \lfloor \mu \cdot t_{\mathrm{OUT}} \rfloor$ \Comment{Target pattern length}
\State Initialize an empty distance map $D[\cdot]$
\For{each training window with start timestamp $t_i-t_{\mathrm{IN}}$ in $\mathcal{D}_{\mathrm{train}}$}
    \State $y_i \gets X_{t_i : t_i + t_{\mathrm{PIN}}}^{V_t}$ \Comment{Ground-truth label segment on $V_t$}
    \State $D[t_i] \gets \left\| y_i - p \right\|_2$ \Comment{Euclidean distance}
\EndFor
\State $k_{\mathrm{poison}} \gets \lceil \alpha_T \cdot |\mathcal{D}_{\mathrm{train}}| \rceil$
\State $k_{\mathrm{sim}} \gets \lfloor \gamma \cdot k_{\mathrm{poison}} \rfloor$
\State $k_{\mathrm{diff}} \gets k_{\mathrm{poison}} - k_{\mathrm{sim}}$
\State $\mathcal{T}_{\mathrm{sim}} \gets \text{TopK}_{t_i}\!\big(-D[t_i],\, k_{\mathrm{sim}}\big)$ \Comment{Most similar}
\State $\mathcal{T}_{\mathrm{diff}} \gets \text{TopK}_{t_i}\!\big(D[t_i],\, k_{\mathrm{diff}}\big)$ \Comment{Most dissimilar}
\State $\mathcal{T}_{\mathrm{poison}} \gets \mathcal{T}_{\mathrm{sim}} \cup \mathcal{T}_{\mathrm{diff}}$
\State \Return $\mathcal{T}_{\mathrm{poison}}$
\end{algorithmic}
\end{algorithm}
In this selection, label-similar samples, whose ground-truths already resemble the target pattern, naturally produce outputs close to it. Consequently, the trigger--target association is easy to fit on them, and relabeling them changes the training labels only slightly. By contrast, label-dissimilar samples naturally suppress the target pattern. When triggered and relabeled, these samples create a conflict: the original features oppose the new target supervision. To resolve this, the model learns the trigger as a high-priority shortcut that overrides the input, thereby strengthening attack effectiveness and robustness.

We formalize our poisoned sample selection as follows. Let $\mathcal{T}_{\mathrm{poison}}=\{t_1, t_2,\ldots, t_k\}$ denote the set of start timestamps of poisoned label windows. Our hybrid selection is defined as:
\begin{equation}
\left\{
\begin{array}{l}
k_{\mathrm{poison}} = \left\lceil \alpha_T \cdot |\mathcal{D}_{\mathrm{train}}| \right\rceil \\
k_{\mathrm{sim}} = \left\lfloor \gamma \cdot k_{\mathrm{poison}} \right\rfloor \\
k_{\mathrm{diff}} = k_{\mathrm{poison}} - k_{\mathrm{sim}} \\
\mathcal{T}_{\mathrm{sim}} = \mathrm{TopK}_{t_i} \left( -d(t_i),\; k_{\mathrm{sim}} \right) \\
\mathcal{T}_{\mathrm{diff}} = \mathrm{TopK}_{t_i} \left( d(t_i),\; k_{\mathrm{diff}} \right) \\
\mathcal{T}_{\mathrm{poison}} = \mathcal{T}_{\mathrm{sim}} \cup \mathcal{T}_{\mathrm{diff}}
\end{array}
\right.,
\end{equation}
where $k_{\mathrm{poison}}$ is the total number of poisoned samples, $k_{\mathrm{sim}}$ and $k_{\mathrm{diff}}$ are the numbers of label-similar and label-dissimilar poisoned samples, respectively, $\mathcal{T}_{\mathrm{sim}}$ and $\mathcal{T}_{\mathrm{diff}}$ denote their timestamp-index sets, $\alpha_{T}$ denotes the poisoning ratio, $\gamma \in [0,1]$ controls the ratio between the two sample types, and $d(t_i)$ is the distance between the label of the sample with start timestamp $t_i-t_{\mathrm{IN}}$ on $V_t$ and the target pattern $p$, where we use the Euclidean distance as the similarity metric:
\begin{equation}
    d(t_i) = \left\| X_{t_i : t_i + t_{\mathrm{PIN}}}^{V_t} - p \right\|_2.
\end{equation}

\begin{callout}{Observation 1}
We find that poisoning only one of the two sample types yields a weaker attack than mixing both, as confirmed by our ablation study in Table~\ref{tab:Ablation}. This is because each type has systematic input features (either encouraging or discouraging the target). When only one type is used, the model attributes the target pattern to these input-specific features rather than the trigger, resulting in a weak and context-dependent trigger--target association.
\end{callout}

\begin{rqbox}
\textbf{Design II:} \emph{Which variables to poison?}
\end{rqbox}

We define the poisoned variables $V_p$ as the input variables into which the trigger is embedded to induce predictions on the target variables $V_t$. Prior works~\cite{lin2024backtime,govindarajulu2023targeted} set $V_p = V_t$, i.e., injecting the trigger directly into the target variables. This choice increases trigger exposure risk for two reasons: (i) abnormal outputs on the target variable naturally prompt scrutiny of its inputs; (ii) concentrating the entire trigger in a single variable yields a longer perturbation on that variable.

We leverage the inherent inter-variable dependencies in multivariate time series to mitigate this. We observe that advanced MLTSF models~\cite{Autoformer,liu2023itransformer,wu2023timesnet,zhou2021informer,pmlr-v162-zhou22g} tend to use inter-variable correlations to enhance their forecasting performance. This means the prediction of the target variables $V_t$ is influenced by their closely correlated variables. Based on this observation, we distribute the trigger across multiple variables that are closely correlated to the target variables $V_t$ rather than embedding it solely in $V_t$. Our ablation study in Table~\ref{tab:Ablation} shows that this selection matters most among the designs of \name, since replacing it with random selection causes the largest degradation of all variants. Specifically, we apply a single-layer GAT to the variable graph, score each variable by its total attention toward $V_t$, and select the top $\lfloor\alpha_P|V|\rfloor$ as $V_p$. The trigger is then split into $|V_p|$ equal fragments of $\lfloor t_{\mathrm{TGR}}/|V_p| \rfloor$ timesteps each and embedded into these variables. This procedure is visualized in Fig.~\ref{GAT} and summarized in Algorithm~\ref{alg:variableselection}.

\begin{figure}[t]
    \centering
    \vspace{0.0cm}
    \includegraphics[width=0.95\linewidth]{figs/GAT.pdf}
    \vspace{0.0cm}
    \caption{Workflow of our poisoned variable selection. From the multivariate series $X$, we build a directed graph and apply a single-layer GAT to obtain edge attentions. Averaged attentions $\bar{\alpha}_{ij}$ are aggregated as $\mathrm{Influence}(i)=\sum_{j\in V_t}\bar{\alpha}_{ij}$, and the top-$k$ variables by this score are selected as $V_p$.}
    \vspace{0.0cm}
    \label{GAT}
\end{figure}

\begin{algorithm}[t]
\caption{Poisoned variable selection}
\label{alg:variableselection}
\begin{algorithmic}[1]
\algrenewcommand\algorithmicrequire{\textbf{Input:}}
\algrenewcommand\algorithmicensure{\textbf{Output:}}
\Require An MLTSF dataset $X\in\mathbb{R}^{T\times N}$, target set $V_t$, ratio $\alpha_P$, heads $n_{\mathrm{head}}$
\Ensure $V_p$
\State $H\in\mathbb{R}^{N\times T} \gets$ standardize $X$
\State $h_i \gets H[i,:]$
\State $V \gets \{1, \dots, N\}$
\State $E \gets V \times V$
\State $G \gets (V, E)$
\State Apply a single-layer GAT with $n_{\mathrm{head}}$ heads to $G$ with node features $H$
\State Compute $\alpha^{(m)}_{ij}$ for all $(i\to j)\in E$ and $m=1,\dots,n_{\mathrm{head}}$ as in Eq.~\eqref{alpha}
\State $\bar{A}[i,j]\gets\bar{\alpha}_{ij}\gets\frac{1}{n_{\mathrm{head}}} \sum_{m=1}^{n_{\mathrm{head}}} \alpha^{(m)}_{ij}$
\For{each $i\in V$}
  \State $\mathrm{Influence}(i)\gets \sum_{j\in V_t}\bar{A}[i,j]$
\EndFor
\State $V_p \gets \mathrm{TopK}\big(\mathrm{Influence}(i),\, k=\lfloor\alpha_P|V|\rfloor\big)$
\State \Return $V_p$
\end{algorithmic}
\end{algorithm}

Given a multivariate time series $X=\{x_1, x_2, \ldots,x_N\} \in \mathbb{R}^{T \times N}$, we represent each variable as a node in a fully connected directed graph $G=(V,E)$. The feature of node $i$ is a standardized time series $h_i=H[i,:] \in \mathbb{R}^T$, where  $H \in \mathbb{R}^{N\times T}$ is constructed by z-scoring each variable over the training horizon. We use a single-layer GAT with $n_{\mathrm{head}}=4$ attention heads, where each head maps a node feature $h_i\in\mathbb{R}^T$ to a scalar, i.e., $W^{(m)}\in\mathbb{R}^{1\times T}$ and $a^{(m)}\in\mathbb{R}^{2}$, so the GAT has only about $4T$ parameters (about 35K on the ETTh1 training split). We adopt GAT because its attention coefficients directly provide a normalized, directed weight of how much each variable contributes to the aggregated representation of every other variable, and they are obtained in a single forward pass over the variable graph. This step takes at most 0.22~s in our experiments, even with 321 variables, which is far cheaper than pretraining the forecasting model to convergence as in the SOTA work~\cite{lin2024backtime}. 
We compute attention on each directed edge $i \rightarrow j$. For head $m \in \{1, \ldots, n_{\mathrm{head}}\}$:
\begin{equation}
\label{alpha}
\begin{aligned}
e_{ij}^{(m)} &= \mathrm{LeakyReLU}\big((a^{(m)})^{\top} [W^{(m)}h_i \,\|\, W^{(m)}h_j]\big),\\
\alpha_{ij}^{(m)} &= \frac{\exp\big(e_{ij}^{(m)}\big)}{\sum_{k \in \mathcal{N}_j} \exp\big(e_{kj}^{(m)}\big)},
\end{aligned}
\end{equation}
where $e_{ij}^{(m)}$ is the unnormalized attention score of edge $i \rightarrow j$, $W^{(m)}$ and $a^{(m)}$ are the parameters of head $m$, $h_i$ is the feature of node $i$, $\|$ denotes vector concatenation, and $\mathcal{N}_j$ is the set of incoming neighbors of node $j$, which equals $V$ since $E=V\times V$. We average $\alpha_{ij}^{(m)}$ over all heads to obtain $\bar{\alpha}_{ij} \in [0,1]$. Node $j$ computes its representation through a softmax-normalized aggregation over its incoming neighbors, and $\bar{\alpha}_{ij}$ reflects the proportion of information from node $i$ in the representation of $j$. We therefore use $\bar{\alpha}_{ij}$ as a measure of the influence of variable $i$ on variable $j$. We then compute a cumulative influence score for each variable $i$ by summing its attention weights toward all target variables:
\begin{equation}
\text{Influence}(i) = \sum_{j \in V_t} \bar{\alpha}_{ij}.
\end{equation}
Based on the influence scores, we select the top $\lfloor\alpha_P|V|\rfloor$ variables as the poisoned set $V_p$.

\begin{callout}{Observation 2}
Selected poisoned variables $V_p$ typically include the target variables, since each target variable exerts a stronger influence on its own prediction than any other variable.
\end{callout}

\begin{rqbox}
\textbf{Design III:} \emph{Where in time to insert triggers?}
\end{rqbox}

After selecting the poisoned variables, we identify effective timestamps for trigger injection on these variables. In time series, strong temporal dependencies and delayed effects make injection time far more critical than spatial trigger placement in images, yet most existing methods ignore this temporal aspect. For example, BackTime~\cite{lin2024backtime} inserts the trigger by overwriting the last $t_{\mathrm{TGR}}$ timesteps of the target variable, based on the intuition that inputs closer to the forecast horizon exert the strongest influence on predictions. This heuristic lacks a data-driven justification, offers limited interpretability, and overlooks that predictive influence between variables is often delayed rather than immediate. For instance, temperature can affect electricity demand several hours later, and changes in heart rate can precede changes in blood pressure. Consequently, a fixed, horizon-adjacent trigger placement may miss the true causal or predictive lags in the data.

To address this, we propose a lag-aware positioning strategy. The complete procedure is summarized in Algorithm~\ref{alg:triggerpositioning}. For each poisoned variable $v_i \in V_p$, we estimate an optimal lag $\delta_{v_i}^*$ that specifies how many steps in advance the trigger should be inserted to most effectively influence predictions on the target variables $V_t$. The trigger fragment for $v_i$ is placed $\delta_{v_i}^*$ steps before the target pattern's onset to ensure it takes effect precisely when the model predicts the target output. Specifically, for each poisoned variable $v_i \in V_p$, we determine its optimal lag $\delta_{v_i}^*$ as follows: 
\begin{equation}
\label{delta}
\begin{aligned}
\delta^*_{v_i} = {}& \argmax_{\delta \in [\lfloor t_{\mathrm{TGR}}/|V_p| \rfloor,\, \delta_{\max}]} \sum_{v_j \in V_t} \sum_{t=0}^{T-t_{\mathrm{IN}}} \\
& \text{Corr}\big(X^{v_i}[t : t+t_{\mathrm{IN}} - \delta],\; X^{v_j}[t+\delta : t+t_{\mathrm{IN}}]\big),
\end{aligned}
\end{equation}
where $\text{Corr}(X^{v_i}, X^{v_j})$ denotes the Pearson correlation coefficient~\cite{benesty2009pearson}:
\begin{equation} 
\text{Corr}(X^{v_i}, X^{v_j}) = \frac{\text{Cov}(X^{v_i}, X^{v_j})}{\sigma_{v_i} \sigma_{v_j}},
\end{equation}
where $\text{Cov}(\cdot,\cdot)$ denotes the covariance, and $\sigma_{v_i}$ and $\sigma_{v_j}$ are the standard deviations of $X^{v_i}$ and $X^{v_j}$, respectively. We search lags up to $\delta_{\max}=50$, so that each correlation in Eq.~\eqref{delta} uses at least $t_{\mathrm{IN}}-\delta_{\max}=46$ observations, and set $\text{Corr}$ to zero when either segment has zero variance. For each candidate lag $\delta$, the objective in Eq.~\eqref{delta} computes the Pearson correlation between the prefix of $v_i$ and the $\delta$-shifted suffix of each target variable $v_j$ within every input window, then aggregates these correlations across all windows and target variables. The optimal lag $\delta^*_{v_i}$ is the one that maximizes this aggregated score, indicating that $v_i$ exhibits the strongest and most consistent association with the targets when leading them by $\delta$ steps. Notably, BackTime's input-end placement is a special case of this strategy: when the strongest association between $v_i$ and $V_t$ is immediate, $\delta^*_{v_i}$ reaches its lower bound and the fragment is placed at the end of the input window. Lag-aware positioning thus generalizes this fixed heuristic and further adapts to delayed dependencies. 

Let $\delta^*=[\delta^*_{v_1}, \ldots, \delta^*_{v_{|V_p|}}]$ denote the optimal lags for all poisoned variables. For a poisoned sample starting at time $t$, the trigger fragment for variable $v_i$ is inserted at the interval $[t + t_{\mathrm{IN}} - \delta^*_{v_i} : t + t_{\mathrm{IN}} - \delta^*_{v_i} + \lfloor t_{\mathrm{TGR}}/|V_p| \rfloor]$. The full trigger is evenly segmented and injected into the poisoned variables according to their respective lags.

\begin{algorithm}
\caption{Trigger temporal positioning}
\label{alg:triggerpositioning}
\begin{algorithmic}[1]
\algrenewcommand\algorithmicrequire{\textbf{Input:}}
\algrenewcommand\algorithmicensure{\textbf{Output:}}
\Require An MLTSF dataset $X\in\mathbb{R}^{T\times N}$, poisoned variables $V_p$, target variables $V_t$, input length $t_{\mathrm{IN}}$, total trigger length $t_{\mathrm{TGR}}$, maximum lag $\delta_{\max}$
\Ensure $\delta^* = [\delta^*_{v_1}, \dots, \delta^*_{v_{|V_p|}}]$
\State $L \gets \lfloor t_{\mathrm{TGR}}/|V_p|\rfloor$ \Comment{minimum feasible lag}
\For{each $v_i \in V_p$}
    \For{$\delta = L$ \textbf{to} $\delta_{\max}$}
        \State $C_{v_i}[\delta] \gets 0$
        \For{$t=0$ \textbf{to} $T - t_{\mathrm{IN}}$}
            \For{each $v_j \in V_t$}
                \State $\mathbf{a} \gets X^{v_i}[t : t{+}t_{\mathrm{IN}}{-}\delta]$
                \State $\mathbf{b} \gets X^{v_j}[t{+}\delta : t{+}t_{\mathrm{IN}}]$
                \State $C_{v_i}[\delta] \mathrel{+}= \mathrm{Corr}(\mathbf{a}, \mathbf{b})$
            \EndFor
        \EndFor
    \EndFor
    \State $\delta^*_{v_i} \gets \argmax_{\delta \in [L,\; \delta_{\max}]} C_{v_i}[\delta]$
\EndFor
\State \Return $\delta^*$
\end{algorithmic}
\end{algorithm}

\begin{rqbox}
\textbf{Design IV:} \emph{What trigger to use?}
\end{rqbox}

To accelerate convergence of both the model and the trigger in subsequent optimization, we construct an effective initial trigger using a data-driven strategy. This strategy selects in-distribution temporal fragments that the model already tends to map to the target pattern under its priors, thereby positioning the optimization closer to a feasible solution. Specifically, we extract trigger values from training samples whose ground-truth labels most closely match the target pattern $p$. The time index $t^*$ is selected by minimizing the Euclidean distance between the label segment and $p$ as follows:
\begin{equation}
    t^* = \argmin_t \left\| X^{V_t}_{t:t+t_{\mathrm{PIN}}} - p \right\|_2.
\end{equation}
For each poisoned variable $v_i \in V_p$, we extract a segment of length $\lfloor t_{\mathrm{TGR}}/|V_p| \rfloor$ by backtracking from $t^*$ according to its optimal lag $\delta^*_{v_i}$:
\begin{equation}
\mathrm{TGR}_{i} = X^{v_i}\big[t^* - \delta_{v_i}^* : t^* - \delta_{v_i}^* + \lfloor t_{\mathrm{TGR}} / |V_p| \rfloor\big].
\end{equation}
The initial trigger matrix $\tau_0$ is then formed by concatenating all trigger segments:
\begin{equation}
    \tau_0 = [\mathrm{TGR}_1, \mathrm{TGR}_2, \ldots, \mathrm{TGR}_{|V_p|}].
\end{equation}
The full procedure is detailed in Algorithm~\ref{alg:triggerinit}.

\begin{algorithm}
\caption{Trigger initialization}
\label{alg:triggerinit}
\begin{algorithmic}[1]
\algrenewcommand\algorithmicrequire{\textbf{Input:}}  %
\algrenewcommand\algorithmicensure{\textbf{Output:}}  %
\Require An MLTSF dataset $X\in\mathbb{R}^{T\times N}$, poisoned variable set $V_p$, target variables $V_t$, target pattern $p$, optimal lags $\delta^*$, target pattern length $t_{\mathrm{PIN}}$, total trigger length $t_{\mathrm{TGR}}$
\Ensure Initial trigger matrix $\tau_0$

\State $t^* \gets \argmin_t \;\big\| X^{V_t}_{t : t+t_{\mathrm{PIN}}} - p \big\|_2$ \Comment{Best match to $p$}
\State Initialize empty trigger matrix $\tau_0$
\For{each poisoned variable $v_i \in V_p$ with lag $\delta^*_{v_i}$}
    \State Extract trigger segment:
    \[
    \mathrm{TGR}_i \gets X^{v_i}\!\left[t^* - \delta^*_{v_i} : \; t^* - \delta^*_{v_i} + \lfloor t_{\mathrm{TGR}}/|V_p| \rfloor\right]
    \]
    \State Append $\mathrm{TGR}_i$ to $\tau_0$
\EndFor
\State \Return $\tau_0 = [\mathrm{TGR}_1, \mathrm{TGR}_2, \dots, \mathrm{TGR}_{|V_p|}]$
\end{algorithmic}
\end{algorithm}

\subsection{Backdoor Training}

Using the four data poisoning designs from Phase I, we now embed the backdoor into the MLTSF model. Specifically, the clean training dataset \(X\) is dynamically poisoned during training according to the poisoning configuration designed in Phase I: samples indexed by $\mathcal{T}_{\mathrm{poison}}$ (Design I) are selected for poisoning, triggers are injected into selected poisoned variables $V_p$ (Design II) at lags $\delta^*$ (Design III), and the initial trigger $\tau_0$ (Design IV) seeds the optimization. These designs are only partially order-dependent: Design I relies only on the target variables and the target pattern and can thus be performed at any point, whereas Designs II--IV must follow this order, since the lags in Design III are estimated for the variables selected in Design II and the initial trigger in Design IV is built from both. This poisoned data is then used to train the model and the trigger via alternating optimization, as detailed in the following subsections. Throughout the whole optimization, $\mathcal{T}_{\mathrm{poison}}$, $V_p$, and $\delta^*$ remain fixed, and the trigger is the only output of Phase I that is updated.

\begin{callout}{Observation 3}
Alternating optimization of the model and the trigger is not a new technique and is widely used in existing backdoor attacks~\cite{huynh2024combat,doan2021lira,nguyen2020input}. However, directly applying this technique to MLTSF leads to a severely weakened attack because stealth constraints yield extremely sparse backdoor signals. Under such sparsity, conventional losses average the few backdoor elements together with the clean segments of poisoned samples, so the model overlooks the trigger–target correlation. Our ablation study (w/o-CustomObj, Table~\ref{tab:Ablation}) shows that this substantially weakens the attack. To overcome sparsity of backdoor signals in MLTSF, we keep the alternating optimization framework but design a series of \textbf{custom loss objectives tailored to the MLTSF task}.
\end{callout}

\subsubsection{Model optimization step}
At model optimization iteration $r$, we fix the current trigger $\tau_k$ and optimize the model parameters $\theta$. When $k=0$, $\tau_k$ is initialized as trigger $\tau_0$ precomputed in Design IV.

Let the current iteration batch be denoted as $B = \{(X_i, Y_i, t_i)\}$, where $t_i$ denotes the forecast start timestamp of sample $i$. We partition $B$ into poisoned and clean subsets using the precomputed timestamp set $\mathcal{T}_{\mathrm{poison}}$:
\begin{equation}
    B_p = \{(X_i, Y_i, t_i) \in B \mid t_i \in \mathcal{T}_{\mathrm{poison}}\}, \; B_c = B \setminus B_p.
\end{equation}

For each \((X_i, Y_i, t_i) \in B_p\) and each poisoned variable \(v_j \in V_p\), we inject the \(j\)-th fragment of \(\tau_k\) into the interval:
\begin{equation}
    \mathcal{I}_j = \Big[t_{\mathrm{IN}} - \delta^*_{v_j} \;:\; t_{\mathrm{IN}} - \delta^*_{v_j} + \lfloor t_{\mathrm{TGR}}/|V_p| \rfloor\Big]
\end{equation} 
via linear interpolation:  
\begin{equation}
    X_i^{v_j}[\mathcal{I}_j] = (1 - \lambda) \, X_i^{v_j}[\mathcal{I}_j] + \lambda \, \tau_k[j],
\end{equation}  
where \(\lambda \in [0,1]\) controls the blend between the original input and the trigger. A smaller $\lambda$ preserves more of the original sequence. We set $\lambda=1$ by default and keep the trigger inconspicuous through the smoothness and range losses introduced below.

Next, we modify the ground-truth label \(Y_i\) as follows. For the target variables \(V_t\), the first \(t_{\mathrm{PIN}} = \lfloor \mu \cdot t_{\mathrm{OUT}} \rfloor\) timesteps are replaced with the target pattern \(p\), while the remaining timesteps and all non-target variables remain unchanged:  
\begin{equation}
\tilde{Y}_i^v = 
\begin{cases}
p, & v \in V_t \text{ at } [0 : t_{\mathrm{PIN}}] \text{ timesteps}  \\
Y_{i}^v, & \text{otherwise}
\end{cases},
\end{equation}
where $t_{\mathrm{PIN}}$ denotes the length of the target pattern and $\mu$ denotes its proportion within the output window. Samples in \(B_c\) are left unmodified.

Then, the batch $B$ containing the poisoned subset $B_p$ and clean subset $B_c$ is fed into the target MLTSF model $f_\theta$ to obtain predictions $\hat{Y}_i=f_\theta(X_i)$, and the total loss is computed using our proposed objective:
\begin{equation}
\label{OptimizationM}
\mathcal{L}_{model} = \alpha_1 \mathcal{L}_c + \alpha_2 \mathcal{L}_p^{V_t} + \alpha_3 \mathcal{L}_p^{\bar{V}_t},
\end{equation}
where $\alpha_1$, $\alpha_2$, and $\alpha_3$ are weights for the three loss terms. In all experiments we fix $\alpha_1=\alpha_3=1$ and leave $\alpha_2$ as the only tuned weight of Eq.~\eqref{OptimizationM}, as it controls how strongly the model is driven toward the target pattern on the target variables of the poisoned samples. Fig.~\ref{Method} provides a visual walkthrough of how these three loss components are computed. The clean loss $\mathcal{L}_c$ is computed over clean samples and ensures that the model maintains its performance on the primary forecasting task:
\begin{equation}
\mathcal{L}_c = \frac{1}{|B_c|} \sum_{i \in B_c} \big( \| \hat{Y}_i - Y_i \|_2^2 + \| \hat{Y}_i - Y_i \|_1 \big).
\end{equation}
Here and below, each term combines the squared $\ell_2$ error and the $\ell_1$ error: the former strongly penalizes large deviations, whereas the latter retains non-vanishing gradients for small residuals and is consistent with the MAE used for evaluation. $\mathcal{L}_p^{V_t}$ is the loss computed on target variables of poisoned samples, enabling the model to learn the mapping between the trigger and target pattern:
\begin{equation}
\mathcal{L}_p^{V_t} = \frac{1}{|B_p|} \sum_{i \in B_p} \big( \| \hat{Y}_{i}^{V_t} - \tilde{Y}^{V_t}_{i} \|_2^2 + \| \hat{Y}_{i}^{V_t} - \tilde{Y}^{V_t}_{i} \|_1 \big),
\end{equation}
where $\tilde{Y}$ denotes poisoned labels with the target pattern. $\mathcal{L}_p^{\bar{V}_t}$ constrains predictions on non-target variables of poisoned samples to match their clean ground-truths, counteracting potential side effects caused by trigger injection into non-target (but poisoned, i.e., $V_p \setminus V_t$) variables and ensuring the attack only affects the target variables:
\begin{equation}
\mathcal{L}_p^{\bar{V}_t} = \frac{1}{|B_p|} \sum_{i \in B_p} \big( \| \hat{Y}_{i}^{\bar{V}_t} - Y_{i}^{\bar{V}_t} \|_2^2 + \| \hat{Y}_{i}^{\bar{V}_t} - Y_{i}^{\bar{V}_t} \|_1 \big).
\end{equation}
Although the three terms share a similar form, we keep them separate to address the extreme sparsity of backdoor signals in MLTSF. In the standard objective in Eq.~\eqref{ConO}, the loss on a poisoned sample is averaged over all $N$ variables and $t_{\mathrm{OUT}}$ steps, so the backdoor supervision on the first $t_{\mathrm{PIN}}$ steps of the $|V_t|$ target variables receives a share of the gradient that is fixed by the data. Our objective instead averages the clean and poisoned losses within their own sample groups and the backdoor and fidelity losses within their own variable groups, so the contribution of the sparse backdoor supervision is set by $\alpha_2$ independently of its frequency in the training data, and each objective can be rebalanced when the poisoning ratio, trigger length, target-pattern length, dataset, or model changes. 

Finally, with learning rate $\eta$, the model parameters are updated via:
\begin{equation}
\theta^{r+1} = \theta^{r} - \eta \cdot \nabla_{\theta^{r}} \mathcal{L}_{model}.
\end{equation}

\subsubsection{Trigger optimization step}
Every $M$ epochs, we freeze the model parameters and update the current trigger $\tau_k$ to adapt it to the evolving model. We iterate over poisoned samples from the training set using the precomputed timestamp set $\mathcal{T}_{\mathrm{poison}}$, embed triggers into them, replace ground-truth labels with the target pattern, as described in the model update step above, after which the poisoned data is fed into the model and the trigger updated by minimizing the loss function:
\begin{equation}
\label{OptimizationT}
\mathcal{L}_{trigger} = \beta_1 \mathcal{L}_p^{V_t} + \beta_2 \mathcal{L}_p^{\bar{V}_t} + \beta_3 \mathcal{L}_{smooth} + \beta_4 \mathcal{L}_{range},
\end{equation}
where $\mathcal{L}_p^{V_t}$ and $\mathcal{L}_p^{\bar{V}_t}$ are identical to those in model optimization. Fig.~\ref{Method} provides a visualization of the computation of $\mathcal{L}_{smooth}$ and $\mathcal{L}_{range}$. The smoothness loss $\mathcal{L}_{smooth}$ penalizes abrupt changes in trigger values to enhance temporal continuity:
\begin{equation}
\begin{aligned}
\mathcal{L}_{smooth} &= \frac{1}{t_{\mathrm{TGR}}} \sum_{i=1}^{\lfloor t_{\mathrm{TGR}}/|V_p| \rfloor} \sum_{j=1}^{|V_p|} \big(s_k^i[j]\big)^2,\\
s_k^i[j] &= \max\big(\big|\tau_k^i[j] - \tau_k^{i-1}[j]\big| - \varepsilon,\, 0\big),
\end{aligned}
\end{equation}
where $\varepsilon$ is the smoothness threshold, set to $0.8$ empirically, and $\tau_k^{0}[j]$ is the input value immediately before the $j$-th fragment, averaged over the batch. The range loss $\mathcal{L}_{range}$ constrains each trigger value within the min–max range of its corresponding poisoned variable:
\begin{equation}
\begin{aligned}
\mathcal{L}_{range} &= \frac{1}{t_{\mathrm{TGR}}} \sum_{i=1}^{\lfloor t_{\mathrm{TGR}}/|V_p| \rfloor} \sum_{j=1}^{|V_p|} r_k^i[j],\\
r_k^i[j] &= \max\big(\min(X^{v_j}) - \tau_k^i[j],\, 0\big)\\
&\quad + \max\big(\tau_k^i[j] - \max(X^{v_j}),\, 0\big).
\end{aligned}
\end{equation}

In Eq.~\eqref{OptimizationT} we likewise fix $\beta_2=1$, $\beta_3=0.01$, and $\beta_4=2$, leaving $\beta_1$ as the only tuned weight. All fixed values are summarized in Table~\ref{symboltable2}.

For a given dataset--model setting, only the two tuned weights $\alpha_2$ and $\beta_1$ are initialized before training and adjusted as needed during training, while the remaining five weights keep the fixed values above. Starting from an initial weight configuration, we train for several epochs and inspect the individual loss curves and validation-sample visualizations. If the objective associated with a loss term remains insufficiently optimized, we increase its corresponding weight and continue training. This process is repeated until the attack effectiveness and clean forecasting performance reach a balanced trade-off. In the default setting (ETTh1 with TimesNet), this procedure yields $\alpha_2=0.4$ and $\beta_1=15$, which are reported as the default values in Table~\ref{symboltable2}. Every other dataset--model pair is tuned in exactly the same way, and no other weight is changed.

\section{Experiments}
\label{experiments}
\subsection{Experimental Setup}
\noindent \textbf{Datasets.} We use seven MLTSF benchmarks: ETTh1~\cite{zhou2021informer}, ETTh2~\cite{zhou2021informer}, ETTm1~\cite{zhou2021informer}, ETTm2~\cite{zhou2021informer}, Electricity~\cite{Autoformer}, Weather~\cite{Autoformer}, and Exchange~\cite{Autoformer}. These datasets cover various real-world scenarios, including power load, meteorological trends, and currency exchange rate fluctuations. Dataset details are provided in Appendix~\ref{Datasets}.

\noindent \textbf{Models.} We evaluate six mainstream MLTSF models spanning MLP-, CNN-, Transformer-, and state-space-based architectures: Autoformer~\cite{Autoformer}, Mamba~\cite{gu2023mamba}, TimesNet~\cite{wu2023timesnet}, iTransformer~\cite{liu2023itransformer}, TSMixer~\cite{chen2023tsmixer}, and TimeMixer~\cite{wang2023timemixer}.

\noindent \textbf{Metrics.} We use Mean Absolute Error (MAE)~\cite{willmott2005advantages} to evaluate forecasting performance and attack effectiveness. Specifically: (i) MAE$_\text{c}$ compares predictions with ground-truth labels under clean inputs; (ii) MAE$_\text{pa}$ compares predictions on $V_t$ under trigger-embedded inputs with the poisoned label, in which the target pattern replaces the first $t_{\mathrm{PIN}}$ steps; (iii) MAE$_\text{pn}$ compares predictions with clean labels on non-target variables under trigger-embedded inputs. Lower values are better for all metrics. For Benign, which has no trigger, the inputs are clean.

\noindent \textbf{Baselines.} We compare \name with two baselines. Benign refers to models trained on clean data and serves as the reference without a backdoor. BackTime~\cite{lin2024backtime} is the SOTA backdoor attack designed specifically for MTSF. Backdoor attacks for time series classification are not directly applicable, since they map a trigger to a class label rather than to a target sequence over a forecast horizon. Their simplest adaptation to forecasting, a fixed trigger as in BadNets~\cite{gu2017badnets}, is evaluated in the ablation study as the RandOnly variant.

\noindent \textbf{Experimental protocol.} Model architectures and training hyperparameters follow the Time Series Library (TSLib)%
\footnote{https://github.com/thuml/Time-Series-Library}. Unless otherwise specified, both input and prediction lengths are 96, with the remaining defaults listed in Table~\ref{symboltable2}. We set the random seed to 2025. In the main experiments, $\lvert V_t\rvert=\max\{1,\operatorname{round}(\alpha_S N)\}$. The variable at zero-based index 1 is always included, and the remaining target variables are sampled uniformly without replacement. The ratios $\alpha_S$ and $\alpha_P$ are scaled per dataset so that every dataset has one target variable and two poisoned variables, each carrying a trigger segment of $\lfloor t_{\mathrm{TGR}}/|V_p| \rfloor=14$ timesteps, i.e., 28 modified timesteps per poisoned input. For each dataset, unless explicitly varied, $V_t$ and the target pattern remain unchanged across models and throughout training. BackTime reuses both from the corresponding \name run.  

\begin{table*}[htbp]
  \centering
  \vspace{0.0cm}
  \caption{Comparison of forecasting ability and attack performance across datasets and models. For each dataset, the three rows report Benign, BackTime~\cite{lin2024backtime}, and \name (shaded). Bold marks the lowest MAE$_\text{pa}$ (i.e., the strongest attack) among them. The TSLib implementation of TSMixer has no instance normalization, hence its higher errors on ETTh2 and ETTm2.}
  \vspace{0.0cm}
  \fontsize{7pt}{7pt}\selectfont
  \renewcommand{\arraystretch}{0.95} %
  \setlength{\tabcolsep}{1.6pt} %
  \resizebox{\linewidth}{!}{
    \begin{tabular}{cc|ccc|ccc|ccc|ccc|ccc|ccc}
    \toprule
    \multirow{2}{*}{Dataset} & \multirow{2}{*}{Method} & \multicolumn{3}{c|}{Autoformer~\cite{Autoformer}} & \multicolumn{3}{c|}{Mamba~\cite{gu2023mamba}} & \multicolumn{3}{c|}{TimesNet~\cite{wu2023timesnet}} & \multicolumn{3}{c|}{iTransformer~\cite{liu2023itransformer}} & \multicolumn{3}{c|}{TSMixer~\cite{chen2023tsmixer}} & \multicolumn{3}{c}{TimeMixer~\cite{wang2023timemixer}} \\
    \cmidrule{3-20}
          &       & MAE$_\text{c}\!\downarrow$ & MAE$_\text{pa}\!\downarrow$ & MAE$_\text{pn}\!\downarrow$ & MAE$_\text{c}\!\downarrow$ & MAE$_\text{pa}\!\downarrow$ & MAE$_\text{pn}\!\downarrow$ & MAE$_\text{c}\!\downarrow$ & MAE$_\text{pa}\!\downarrow$ & MAE$_\text{pn}\!\downarrow$ & MAE$_\text{c}\!\downarrow$ & MAE$_\text{pa}\!\downarrow$ & MAE$_\text{pn}\!\downarrow$ & MAE$_\text{c}\!\downarrow$ & MAE$_\text{pa}\!\downarrow$ & MAE$_\text{pn}\!\downarrow$ & MAE$_\text{c}\!\downarrow$ & MAE$_\text{pa}\!\downarrow$ & MAE$_\text{pn}\!\downarrow$ \\
    \midrule
    \multirow{3}{*}{ETTh1} & Benign & 0.439 & 0.801 & 0.444 & 0.445 & 0.782 & 0.455 & 0.424 & 0.781 & 0.432 & 0.411 & 0.776 & 0.421 & 0.503 & 1.009 & 0.507 & 0.393 & 0.768 & 0.401 \\
     & BackTime & 0.565 & 0.817 & 0.522 & 0.456 & 0.776 & 0.460 & 0.448 & 0.750 & 0.439 & 0.406 & 0.769 & 0.415 & 0.478 & 0.547 & 0.480 & 0.417 & 0.641 & 0.415 \\
     & \cellcolor{RQFrame!12}\name & \cellcolor{RQFrame!12}0.465 & \cellcolor{RQFrame!12}\textbf{0.485} & \cellcolor{RQFrame!12}0.460 & \cellcolor{RQFrame!12}0.477 & \cellcolor{RQFrame!12}\textbf{0.478} & \cellcolor{RQFrame!12}0.459 & \cellcolor{RQFrame!12}0.435 & \cellcolor{RQFrame!12}\textbf{0.336} & \cellcolor{RQFrame!12}0.440 & \cellcolor{RQFrame!12}0.406 & \cellcolor{RQFrame!12}\textbf{0.234} & \cellcolor{RQFrame!12}0.414 & \cellcolor{RQFrame!12}0.509 & \cellcolor{RQFrame!12}\textbf{0.372} & \cellcolor{RQFrame!12}0.526 & \cellcolor{RQFrame!12}0.413 & \cellcolor{RQFrame!12}\textbf{0.255} & \cellcolor{RQFrame!12}0.415 \\
    \midrule
    \multirow{3}{*}{ETTh2} & Benign & 0.395 & 0.782 & 0.409 & 0.372 & 0.790 & 0.385 & 0.374 & 0.793 & 0.387 & 0.349 & 0.784 & 0.362 & 0.874 & 0.961 & 0.956 & 0.342 & 0.792 & 0.354 \\
     & BackTime & 0.456 & 0.763 & 0.454 & 0.397 & 0.811 & 0.410 & 0.380 & 0.792 & 0.384 & 0.352 & 0.758 & 0.365 & 1.021 & 0.717 & 1.136 & 0.363 & 0.694 & 0.370 \\
     & \cellcolor{RQFrame!12}\name & \cellcolor{RQFrame!12}0.417 & \cellcolor{RQFrame!12}\textbf{0.412} & \cellcolor{RQFrame!12}0.442 & \cellcolor{RQFrame!12}0.406 & \cellcolor{RQFrame!12}\textbf{0.391} & \cellcolor{RQFrame!12}0.428 & \cellcolor{RQFrame!12}0.361 & \cellcolor{RQFrame!12}\textbf{0.329} & \cellcolor{RQFrame!12}0.355 & \cellcolor{RQFrame!12}0.366 & \cellcolor{RQFrame!12}\textbf{0.216} & \cellcolor{RQFrame!12}0.379 & \cellcolor{RQFrame!12}0.769 & \cellcolor{RQFrame!12}\textbf{0.617} & \cellcolor{RQFrame!12}0.886 & \cellcolor{RQFrame!12}0.363 & \cellcolor{RQFrame!12}\textbf{0.274} & \cellcolor{RQFrame!12}0.377 \\
    \midrule
    \multirow{3}{*}{ETTm1} & Benign & 0.508 & 0.796 & 0.535 & 0.386 & 0.779 & 0.396 & 0.376 & 0.777 & 0.387 & 0.367 & 0.793 & 0.379 & 0.445 & 0.885 & 0.460 & 0.366 & 0.782 & 0.378 \\
     & BackTime & 0.482 & 0.881 & 0.458 & 0.414 & 0.781 & 0.421 & 0.381 & 0.819 & 0.388 & 0.366 & 0.801 & 0.376 & 0.448 & 0.736 & 0.449 & 0.373 & 0.484 & 0.366 \\
     & \cellcolor{RQFrame!12}\name & \cellcolor{RQFrame!12}0.525 & \cellcolor{RQFrame!12}\textbf{0.501} & \cellcolor{RQFrame!12}0.532 & \cellcolor{RQFrame!12}0.386 & \cellcolor{RQFrame!12}\textbf{0.399} & \cellcolor{RQFrame!12}0.395 & \cellcolor{RQFrame!12}0.384 & \cellcolor{RQFrame!12}\textbf{0.355} & \cellcolor{RQFrame!12}0.393 & \cellcolor{RQFrame!12}0.356 & \cellcolor{RQFrame!12}\textbf{0.186} & \cellcolor{RQFrame!12}0.379 & \cellcolor{RQFrame!12}0.492 & \cellcolor{RQFrame!12}\textbf{0.192} & \cellcolor{RQFrame!12}0.505 & \cellcolor{RQFrame!12}0.373 & \cellcolor{RQFrame!12}\textbf{0.372} & \cellcolor{RQFrame!12}0.388 \\
    \midrule
    \multirow{3}{*}{ETTm2} & Benign & 0.295 & 0.512 & 0.306 & 0.275 & 0.516 & 0.283 & 0.268 & 0.517 & 0.276 & 0.274 & 0.517 & 0.280 & 0.479 & 0.522 & 0.475 & 0.261 & 0.515 & 0.267 \\
     & BackTime & 0.347 & 0.643 & 0.338 & 0.278 & 0.532 & 0.283 & 0.269 & 0.526 & 0.274 & 0.266 & 0.552 & 0.273 & 0.418 & 0.464 & 0.420 & 0.262 & 0.537 & 0.265 \\
     & \cellcolor{RQFrame!12}\name & \cellcolor{RQFrame!12}0.297 & \cellcolor{RQFrame!12}\textbf{0.315} & \cellcolor{RQFrame!12}0.322 & \cellcolor{RQFrame!12}0.281 & \cellcolor{RQFrame!12}\textbf{0.281} & \cellcolor{RQFrame!12}0.306 & \cellcolor{RQFrame!12}0.267 & \cellcolor{RQFrame!12}\textbf{0.223} & \cellcolor{RQFrame!12}0.271 & \cellcolor{RQFrame!12}0.261 & \cellcolor{RQFrame!12}\textbf{0.203} & \cellcolor{RQFrame!12}0.284 & \cellcolor{RQFrame!12}0.374 & \cellcolor{RQFrame!12}\textbf{0.324} & \cellcolor{RQFrame!12}0.430 & \cellcolor{RQFrame!12}0.266 & \cellcolor{RQFrame!12}\textbf{0.169} & \cellcolor{RQFrame!12}0.274 \\
    \midrule
    \multirow{3}{*}{Electricity} & Benign & 0.323 & 0.854 & 0.347 & 0.317 & 0.761 & 0.342 & 0.285 & 0.732 & 0.305 & 0.245 & 0.725 & 0.288 & 0.312 & 0.814 & 0.359 & 0.253 & 0.718 & 0.283 \\
     & BackTime & 0.347 & 0.875 & 0.353 & 0.336 & 0.753 & 0.361 & 0.305 & 0.764 & 0.299 & 0.268 & 0.711 & 0.346 & 0.339 & 0.738 & 0.366 & 0.271 & 0.627 & 0.295 \\
     & \cellcolor{RQFrame!12}\name & \cellcolor{RQFrame!12}0.335 & \cellcolor{RQFrame!12}\textbf{0.344} & \cellcolor{RQFrame!12}0.330 & \cellcolor{RQFrame!12}0.324 & \cellcolor{RQFrame!12}\textbf{0.317} & \cellcolor{RQFrame!12}0.325 & \cellcolor{RQFrame!12}0.298 & \cellcolor{RQFrame!12}\textbf{0.206} & \cellcolor{RQFrame!12}0.299 & \cellcolor{RQFrame!12}0.283 & \cellcolor{RQFrame!12}\textbf{0.168} & \cellcolor{RQFrame!12}0.291 & \cellcolor{RQFrame!12}0.322 & \cellcolor{RQFrame!12}\textbf{0.276} & \cellcolor{RQFrame!12}0.344 & \cellcolor{RQFrame!12}0.249 & \cellcolor{RQFrame!12}\textbf{0.239} & \cellcolor{RQFrame!12}0.275 \\
    \midrule
    \multirow{3}{*}{Weather} & Benign & 0.364 & 0.827 & 0.364 & 0.252 & 0.799 & 0.252 & 0.245 & 0.795 & 0.245 & 0.227 & 0.778 & 0.227 & 0.256 & 0.816 & 0.256 & 0.210 & 0.791 & 0.209 \\
     & BackTime & 0.325 & 0.800 & 0.324 & 0.246 & 0.777 & 0.242 & 0.221 & 0.726 & 0.216 & 0.213 & 0.724 & 0.213 & 0.240 & 0.596 & 0.233 & 0.210 & 0.535 & 0.206 \\
     & \cellcolor{RQFrame!12}\name & \cellcolor{RQFrame!12}0.389 & \cellcolor{RQFrame!12}\textbf{0.447} & \cellcolor{RQFrame!12}0.382 & \cellcolor{RQFrame!12}0.245 & \cellcolor{RQFrame!12}\textbf{0.231} & \cellcolor{RQFrame!12}0.242 & \cellcolor{RQFrame!12}0.233 & \cellcolor{RQFrame!12}\textbf{0.202} & \cellcolor{RQFrame!12}0.238 & \cellcolor{RQFrame!12}0.225 & \cellcolor{RQFrame!12}\textbf{0.209} & \cellcolor{RQFrame!12}0.224 & \cellcolor{RQFrame!12}0.238 & \cellcolor{RQFrame!12}\textbf{0.172} & \cellcolor{RQFrame!12}0.231 & \cellcolor{RQFrame!12}0.212 & \cellcolor{RQFrame!12}\textbf{0.179} & \cellcolor{RQFrame!12}0.213 \\
    \midrule
    \multirow{3}{*}{Exchange} & Benign & 0.325 & 0.634 & 0.352 & 0.271 & 0.544 & 0.299 & 0.239 & 0.531 & 0.261 & 0.221 & 0.542 & 0.242 & 0.334 & 0.507 & 0.362 & 0.202 & 0.537 & 0.222 \\
     & BackTime & 0.287 & 0.540 & 0.316 & 0.263 & 0.543 & 0.290 & 0.233 & 0.558 & 0.254 & 0.204 & 0.505 & 0.223 & 0.401 & 0.463 & 0.396 & 0.213 & 0.440 & 0.233 \\
     & \cellcolor{RQFrame!12}\name & \cellcolor{RQFrame!12}0.298 & \cellcolor{RQFrame!12}\textbf{0.293} & \cellcolor{RQFrame!12}0.329 & \cellcolor{RQFrame!12}0.286 & \cellcolor{RQFrame!12}\textbf{0.228} & \cellcolor{RQFrame!12}0.326 & \cellcolor{RQFrame!12}0.238 & \cellcolor{RQFrame!12}\textbf{0.204} & \cellcolor{RQFrame!12}0.245 & \cellcolor{RQFrame!12}0.218 & \cellcolor{RQFrame!12}\textbf{0.106} & \cellcolor{RQFrame!12}0.278 & \cellcolor{RQFrame!12}0.347 & \cellcolor{RQFrame!12}\textbf{0.175} & \cellcolor{RQFrame!12}0.411 & \cellcolor{RQFrame!12}0.211 & \cellcolor{RQFrame!12}\textbf{0.169} & \cellcolor{RQFrame!12}0.249 \\
    \midrule
    \multirow{3}{*}{Average} & Benign & 0.378 & 0.744 & 0.394 & 0.331 & 0.710 & 0.345 & 0.316 & 0.704 & 0.328 & 0.299 & 0.702 & 0.314 & 0.458 & 0.788 & 0.482 & 0.290 & 0.700 & 0.302 \\
     & BackTime & 0.401 & 0.760 & 0.395 & 0.341 & 0.710 & 0.352 & 0.320 & 0.705 & 0.322 & 0.296 & 0.689 & 0.316 & 0.478 & 0.609 & 0.497 & 0.301 & 0.565 & 0.307 \\
     & \cellcolor{RQFrame!12}\name & \cellcolor{RQFrame!12}0.389 & \cellcolor{RQFrame!12}\textbf{0.400} & \cellcolor{RQFrame!12}0.400 & \cellcolor{RQFrame!12}0.344 & \cellcolor{RQFrame!12}\textbf{0.332} & \cellcolor{RQFrame!12}0.354 & \cellcolor{RQFrame!12}0.317 & \cellcolor{RQFrame!12}\textbf{0.265} & \cellcolor{RQFrame!12}0.320 & \cellcolor{RQFrame!12}0.302 & \cellcolor{RQFrame!12}\textbf{0.189} & \cellcolor{RQFrame!12}0.321 & \cellcolor{RQFrame!12}0.436 & \cellcolor{RQFrame!12}\textbf{0.304} & \cellcolor{RQFrame!12}0.476 & \cellcolor{RQFrame!12}0.298 & \cellcolor{RQFrame!12}\textbf{0.237} & \cellcolor{RQFrame!12}0.313 \\
    \bottomrule
    \end{tabular}%
    }
    \vspace{0.0cm}
  \label{tab:comparison1}%
\end{table*}%

\subsection{Main Results}

\begin{figure*}[t]
    \centering
    \vspace{0.0cm}
    \includegraphics[width=0.96\textwidth]{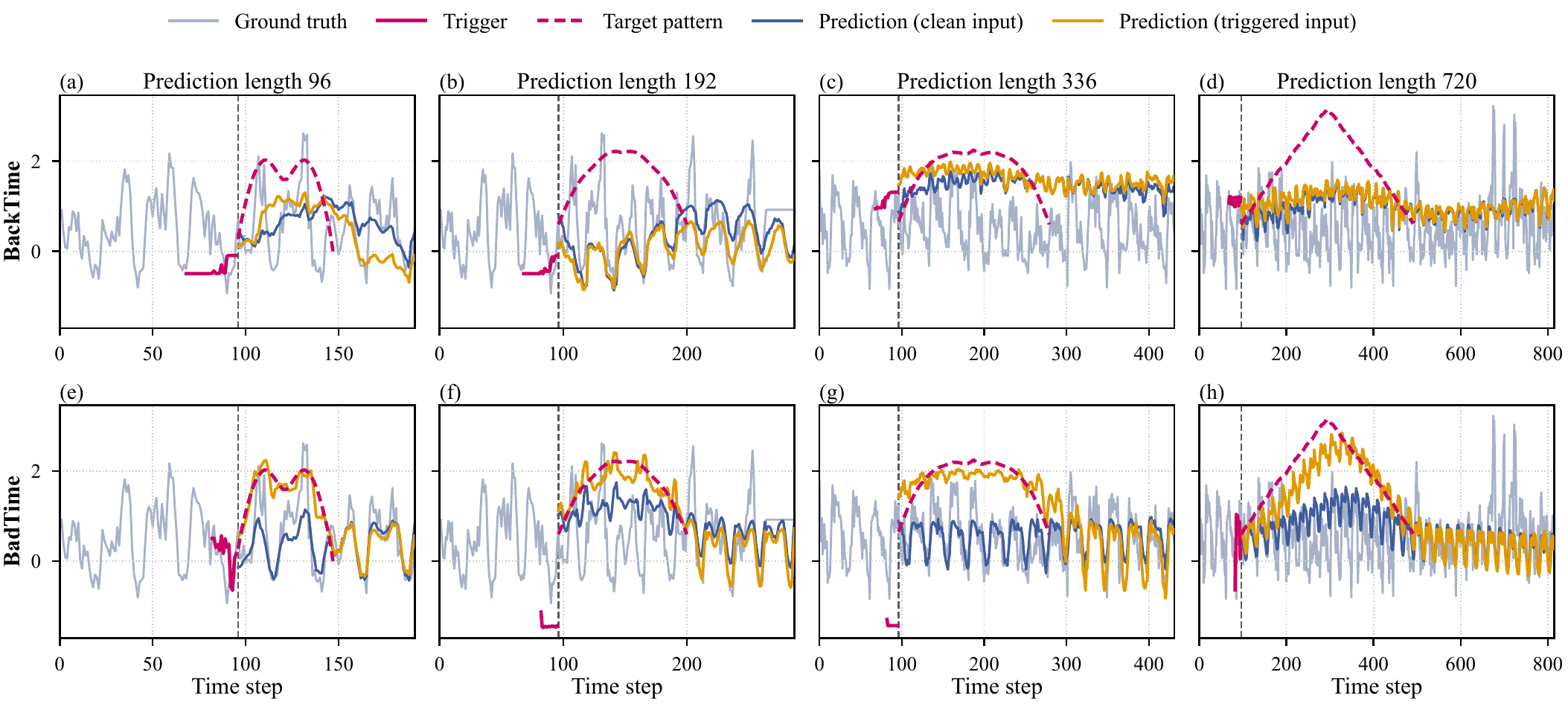}
    \vspace{0.0cm}
    \caption{Attack examples of BackTime (top row, (a)--(d)) and \name (bottom row, (e)--(h)) on the same ETTh1 window, showing the target variable under prediction lengths 96, 192, 336, and 720. The dashed vertical line marks the start of the forecast window.}
    \vspace{0.0cm}
    \label{badtime_vs_backtime}
\end{figure*}
\subsubsection{Effectiveness comparison}
 Table~\ref{tab:comparison1} compares Benign, BackTime, and \name on six mainstream models and seven benchmark datasets.  \name achieves the lowest MAE$_\text{pa}$ in all 42 dataset--model settings. Averaged over these settings, it lowers MAE$_\text{pa}$ from 0.673 under BackTime to 0.288, a reduction of 57.2\%. This attack strength comes at almost no cost to forecasting: relative to Benign, the average MAE$_\text{c}$ and MAE$_\text{pn}$ of \name increase by only 0.7\% and 0.9\%, and neither metric increases by more than 0.05 in any setting. The small change in MAE$_\text{pn}$ shows that the trigger has almost no side effect on the non-target variables, because the fidelity loss $\mathcal{L}_p^{\bar{V}_t}$ keeps their predictions on poisoned samples close to the clean labels. Fig.~\ref{badtime_vs_backtime} provides a visual attack case of BackTime and \name on the same time series window from the ETTh1 dataset with the magenta curves denoting the embedded trigger (solid) and the target pattern (dashed). For clarity, only the target variable is shown. As illustrated, \name produces poisoned predictions that better match the target pattern than those of BackTime, with a trigger that is shorter, smoother, and more natural than that of BackTime. \name also better preserves forecasting performance on clean inputs, with predictions more closely aligned with ground-truth labels compared to BackTime.

\subsubsection{Robustness comparison to longer forecast lengths}
To assess the robustness of \name to longer forecast lengths, we conduct experiments on the ETTh1 dataset using the TimesNet model, gradually extending the prediction horizon from 96 to 192, 336, and 720, with results in Table~\ref{varypredlength}. At each prediction length, the target pattern is a Hump of length $t_{\mathrm{PIN}}=\lfloor\mu\cdot t_{\mathrm{OUT}}\rfloor$. Both BackTime and \name maintain forecasting performance comparable to Benign at all four lengths. \name remains effective even at 720 steps and reduces MAE$_\text{pa}$ by about 40\%--55\% compared to BackTime. In contrast, BackTime lowers MAE$_\text{pa}$ only slightly from that of Benign at every length, showing that its trigger barely steers long-horizon predictions.
\begin{table}[t]
  \centering
  \fontsize{14pt}{14pt}\selectfont
  \setlength{\tabcolsep}{2pt} %
  \caption{Forecasting performance and attack effectiveness of Benign, BackTime, and \name under varying prediction lengths. Bold marks the lowest MAE$_\text{pa}$ (i.e., the strongest attack) among them.}
  \resizebox{\linewidth}{!}{
    \begin{tabular}{c|cc>{\columncolor{RQFrame!12}}c|cc>{\columncolor{RQFrame!12}}c|cc>{\columncolor{RQFrame!12}}c}
    \toprule
    \multirow{2}[4]{*}{Horizon} & \multicolumn{3}{c|}{MAE$_\text{c}\!{\downarrow}$} & \multicolumn{3}{c|}{MAE$_\text{pa}\!{\downarrow}$} & \multicolumn{3}{c}{MAE$_\text{pn}\!{\downarrow}$} \\
\cmidrule{2-10}          & Benign & BackTime & \name & Benign & BackTime & \name & Benign & BackTime & \name \\
    \midrule
    96    & 0.424 & 0.448 & 0.435 & 0.781 & 0.750 & \textbf{0.336} & 0.432 & 0.439 & 0.440 \\ %
    192   & 0.464 & 0.474 & 0.459 & 0.848 & 0.768 & \textbf{0.465} & 0.476 & 0.455 & 0.468 \\
    336   & 0.470 & 0.512 & 0.474 & 0.904 & 0.781 & \textbf{0.467} & 0.484 & 0.480 & 0.479 \\
    720   & 0.493 & 0.510 & 0.508 & 0.960 & 0.857 & \textbf{0.505} & 0.509 & 0.501 & 0.516 \\
    \bottomrule
    \end{tabular}%
    }
  \label{varypredlength}%
\end{table}%

\subsubsection{Stealthiness comparison}
We evaluate the detectability of \name using USAD~\cite{audibert2020usad} and GDN~\cite{deng2021graph}, the two anomaly detectors used in the evaluation of BackTime~\cite{lin2024backtime}. Both detectors are trained on clean data preceding the test split. We report the ROC-AUC and the true positive rate (TPR) at a false positive rate (FPR) of 5\%, whose chance levels are 0.5 and 0.05. Following the threat model, each attack is inspected on the data it exposes to the victim: BackTime on test series poisoned by its injection procedure and \name on its triggered inputs. We also include a control that applies a three-standard-deviation shift away from the training mean at the positions of the \name trigger.

Table~\ref{tab:comparsion2} shows that the trigger of \name passes as normal data. Averaged over the six forecasting models, the AUC on its triggered inputs stays between 0.44 and 0.58 under USAD and between 0.50 and 0.63 under GDN, and the TPR at 5\% FPR never exceeds 0.08. Both detectors therefore raise alarms on triggered windows about as often as on clean ones. The detectors themselves are sensitive, since the control is detected with AUCs from 0.63 to 1.00. What escapes them is the trigger, which is initialized from real segments of the data and bounded by the smoothness and range losses, so that the detectors find little in it that departs from normal data. In every average row of Table~\ref{tab:comparsion2}, the triggered inputs of \name also lie closer to chance level than the series poisoned by BackTime. On ETTh1, ETTh2, and ETTm1, both detectors separate the poisoned series of BackTime from clean data with AUCs between 0.60 and 0.80, while the triggered inputs of \name remain close to chance. The largest departure from chance for \name is GDN on ETTm2 with an AUC of 0.632, and even there the detector flags no more triggered windows than clean ones at the 5\% operating point.
\begin{table*}[t]
  \centering
  \vspace{0.0cm}
  \caption{Stealthiness of BackTime~\cite{lin2024backtime} and \name against the anomaly detectors USAD~\cite{audibert2020usad} and GDN~\cite{deng2021graph}. BackTime is inspected on test series poisoned by its own injection procedure and \name on its triggered inputs. TPR is measured at a 5\% false positive rate, so chance level is 0.5 for AUC and 0.05 for TPR. The control adds a shift of three standard deviations, away from the training mean, at the positions of the \name trigger. In the average rows, bold marks the attack whose value is closer to chance level.}
  \vspace{0.0cm}
  \fontsize{16pt}{16pt}\selectfont
  \renewcommand{\arraystretch}{0.95}\setlength{\tabcolsep}{14pt} %
  \resizebox{\linewidth}{!}{
    \begin{tabular}{c|c|c|cc|cc|cc|cc|cc|cc}
    \toprule
    \multicolumn{1}{c}{\multirow{2}[4]{*}{\makecell{Anomaly\\Detection}}} & \multicolumn{1}{c}{\multirow{2}[4]{*}{Models}} & \multicolumn{1}{c|}{\multirow{2}[4]{*}{\makecell{Backdoor\\Attacks}}} & \multicolumn{2}{c|}{ETTh1} & \multicolumn{2}{c|}{ETTh2} & \multicolumn{2}{c|}{ETTm1} & \multicolumn{2}{c|}{ETTm2} & \multicolumn{2}{c|}{Weather} & \multicolumn{2}{c}{Exchange} \\
    \cmidrule{4-15}    \multicolumn{1}{c}{} & \multicolumn{1}{c}{} & & AUC & TPR & AUC & TPR & AUC & TPR & AUC & TPR & AUC & TPR & AUC & TPR \\
    \midrule
    \multirow{15}{*}{USAD} & \multirow{2}{*}{Autoformer} & BackTime & 0.660 & 0.096 & 0.699 & 0.070 & 0.638 & 0.093 & 0.426 & 0.056 & 0.535 & 0.068 & 0.467 & 0.000 \\
     &  & \name & 0.499 & 0.051 & 0.490 & 0.046 & 0.580 & 0.079 & 0.487 & 0.051 & 0.538 & 0.054 & 0.487 & 0.057 \\
     & \multirow{2}{*}{Mamba} & BackTime & 0.614 & 0.069 & 0.695 & 0.087 & 0.624 & 0.104 & 0.443 & 0.043 & 0.559 & 0.075 & 0.438 & 0.000 \\
     &  & \name & 0.502 & 0.052 & 0.470 & 0.044 & 0.572 & 0.077 & 0.417 & 0.052 & 0.532 & 0.052 & 0.488 & 0.056 \\
     & \multirow{2}{*}{TimesNet} & BackTime & 0.605 & 0.104 & 0.704 & 0.105 & 0.613 & 0.086 & 0.431 & 0.047 & 0.539 & 0.050 & 0.470 & 0.000 \\
     &  & \name & 0.505 & 0.052 & 0.480 & 0.045 & 0.572 & 0.077 & 0.430 & 0.044 & 0.548 & 0.068 & 0.488 & 0.057 \\
     & \multirow{2}{*}{iTransformer} & BackTime & 0.668 & 0.117 & 0.713 & 0.104 & 0.648 & 0.124 & 0.429 & 0.051 & 0.531 & 0.067 & 0.448 & 0.000 \\
     &  & \name & 0.507 & 0.053 & 0.495 & 0.046 & 0.596 & 0.085 & 0.437 & 0.048 & 0.548 & 0.068 & 0.485 & 0.056 \\
     & \multirow{2}{*}{TSMixer} & BackTime & 0.635 & 0.086 & 0.704 & 0.080 & 0.621 & 0.116 & 0.439 & 0.041 & 0.542 & 0.054 & 0.463 & 0.000 \\
     &  & \name & 0.505 & 0.052 & 0.480 & 0.045 & 0.586 & 0.081 & 0.433 & 0.049 & 0.542 & 0.064 & 0.488 & 0.057 \\
     & \multirow{2}{*}{TimeMixer} & BackTime & 0.644 & 0.115 & 0.689 & 0.065 & 0.637 & 0.126 & 0.439 & 0.050 & 0.541 & 0.053 & 0.448 & 0.000 \\
     &  & \name & 0.511 & 0.054 & 0.476 & 0.045 & 0.581 & 0.080 & 0.435 & 0.049 & 0.526 & 0.055 & 0.485 & 0.056 \\
    \cmidrule{2-15}
     & \multirow{2}{*}{Average} & BackTime & 0.638 & 0.098 & 0.701 & 0.085 & 0.630 & 0.108 & 0.434 & 0.048 & 0.541 & 0.061 & 0.456 & 0.000 \\
     &  & \cellcolor{RQFrame!12}\name & \cellcolor{RQFrame!12}\textbf{0.505} & \cellcolor{RQFrame!12}\textbf{0.052} & \cellcolor{RQFrame!12}\textbf{0.482} & \cellcolor{RQFrame!12}\textbf{0.045} & \cellcolor{RQFrame!12}\textbf{0.581} & \cellcolor{RQFrame!12}\textbf{0.080} & \cellcolor{RQFrame!12}\textbf{0.440} & \cellcolor{RQFrame!12}\textbf{0.049} & \cellcolor{RQFrame!12}\textbf{0.539} & \cellcolor{RQFrame!12}\textbf{0.060} & \cellcolor{RQFrame!12}\textbf{0.487} & \cellcolor{RQFrame!12}\textbf{0.056} \\
    \cmidrule{2-15}
     & \multicolumn{2}{c|}{Control} & 0.884 & 0.382 & 0.941 & 0.461 & 0.762 & 0.254 & 0.921 & 0.447 & 0.633 & 0.205 & 0.835 & 0.573 \\
    \midrule
    \multirow{15}{*}{GDN} & \multirow{2}{*}{Autoformer} & BackTime & 0.636 & 0.074 & 0.811 & 0.083 & 0.642 & 0.127 & 0.650 & 0.046 & 0.571 & 0.041 & 0.573 & 0.000 \\
     &  & \name & 0.500 & 0.049 & 0.559 & 0.048 & 0.505 & 0.051 & 0.624 & 0.043 & 0.517 & 0.050 & 0.464 & 0.082 \\
     & \multirow{2}{*}{Mamba} & BackTime & 0.595 & 0.050 & 0.798 & 0.092 & 0.633 & 0.124 & 0.638 & 0.025 & 0.575 & 0.053 & 0.624 & 0.000 \\
     &  & \name & 0.500 & 0.050 & 0.501 & 0.048 & 0.504 & 0.051 & 0.632 & 0.043 & 0.576 & 0.050 & 0.561 & 0.097 \\
     & \multirow{2}{*}{TimesNet} & BackTime & 0.557 & 0.058 & 0.803 & 0.042 & 0.661 & 0.111 & 0.638 & 0.027 & 0.577 & 0.050 & 0.630 & 0.001 \\
     &  & \name & 0.502 & 0.050 & 0.527 & 0.048 & 0.504 & 0.051 & 0.618 & 0.043 & 0.559 & 0.050 & 0.464 & 0.011 \\
     & \multirow{2}{*}{iTransformer} & BackTime & 0.574 & 0.035 & 0.796 & 0.043 & 0.653 & 0.175 & 0.642 & 0.018 & 0.584 & 0.061 & 0.554 & 0.000 \\
     &  & \name & 0.502 & 0.050 & 0.576 & 0.048 & 0.511 & 0.051 & 0.670 & 0.043 & 0.559 & 0.050 & 0.560 & 0.079 \\
     & \multirow{2}{*}{TSMixer} & BackTime & 0.584 & 0.032 & 0.804 & 0.061 & 0.629 & 0.104 & 0.642 & 0.018 & 0.570 & 0.042 & 0.577 & 0.000 \\
     &  & \name & 0.502 & 0.051 & 0.527 & 0.048 & 0.515 & 0.051 & 0.630 & 0.042 & 0.572 & 0.047 & 0.464 & 0.011 \\
     & \multirow{2}{*}{TimeMixer} & BackTime & 0.636 & 0.077 & 0.792 & 0.026 & 0.640 & 0.120 & 0.638 & 0.041 & 0.586 & 0.042 & 0.553 & 0.000 \\
     &  & \name & 0.504 & 0.050 & 0.571 & 0.048 & 0.513 & 0.051 & 0.620 & 0.043 & 0.544 & 0.048 & 0.573 & 0.066 \\
    \cmidrule{2-15}
     & \multirow{2}{*}{Average} & BackTime & 0.597 & 0.054 & 0.801 & 0.058 & 0.643 & 0.127 & 0.641 & 0.029 & 0.577 & 0.048 & 0.585 & 0.000 \\
     &  & \cellcolor{RQFrame!12}\name & \cellcolor{RQFrame!12}\textbf{0.502} & \cellcolor{RQFrame!12}\textbf{0.050} & \cellcolor{RQFrame!12}\textbf{0.543} & \cellcolor{RQFrame!12}\textbf{0.048} & \cellcolor{RQFrame!12}\textbf{0.509} & \cellcolor{RQFrame!12}\textbf{0.051} & \cellcolor{RQFrame!12}\textbf{0.632} & \cellcolor{RQFrame!12}\textbf{0.043} & \cellcolor{RQFrame!12}\textbf{0.555} & \cellcolor{RQFrame!12}\textbf{0.049} & \cellcolor{RQFrame!12}\textbf{0.514} & \cellcolor{RQFrame!12}\textbf{0.058} \\
    \cmidrule{2-15}
     & \multicolumn{2}{c|}{Control} & 0.976 & 0.858 & 0.999 & 0.999 & 0.742 & 0.272 & 1.000 & 1.000 & 0.909 & 0.666 & 0.940 & 0.044 \\
    \bottomrule
    \end{tabular}%
  }
  \vspace{0.0cm}
  \label{tab:comparsion2}%
\end{table*}%

\subsection{Resistance to Model-Level Defenses}\label{sec:defense}
 The detectors above inspect the data. We now examine whether the victim can remove the backdoor from the model itself. Since the adversary controls training, the victim can apply such a defense only after receiving the model and before deploying it. The victim holds its own training data, but having outsourced training to save computation, it can afford to use only part of the data for defense. We vary this part from 5\% to 50\% of the training data and, following~\cite{liu2018fine}, evaluate three defenses that require no knowledge of the trigger or the target pattern:
\begin{enumerate}
    \item \textbf{Pruning:} It removes the neurons that stay dormant on clean inputs, where the backdoor is assumed to reside. We rank the hidden channels of the last TimesBlock by their mean activation on the clean data and prune the 90\% least active ones.
    \item \textbf{Fine-tuning:} It retrains the backdoored model on the clean data so that the model gradually forgets the backdoor. We fine-tune all parameters for 10 epochs.
    \item \textbf{Fine-pruning:} It fine-tunes the pruned model on the clean data, combining the two defenses above with the same settings.
\end{enumerate}
\noindent Each defense uses the same clean data, sampled with the three random seeds 2025, 2026, and 2027. Experiments use TimesNet on ETTh1, ETTm1, and Weather. We measure each defense by its removal rate
\begin{equation*}
\text{Removal rate}=\frac{\text{MAE}_\text{pa}^\text{def}-\text{MAE}_\text{pa}^\text{bd}}{\text{MAE}_\text{pa}^\text{ben}-\text{MAE}_\text{pa}^\text{bd}},
\end{equation*}
\noindent where the superscripts denote the defended, backdoored, and benign models. The rate is 0\% when the backdoor is intact and 100\% when the defended model deviates from the target pattern as much as the benign model.

Table~\ref{tab:defense} reports the removal rates. Pruning has a limited effect even though it removes 90\% of the channels. The trigger of \name is initialized from in-distribution segments of real data and optimized under the smoothness and range losses, so it is processed by the channels that are also active on clean inputs, which pruning keeps. Fine-tuning removes more of the backdoor as the clean data grows. Fine-pruning improves on fine-tuning on ETTh1 and ETTm1, whereas on Weather it falls behind fine-tuning once the defender has 10\% or more of the data. With 5\% or 10\% of the training data, a considerable part of the backdoor survives all three defenses. Among the four budgets evaluated, only the largest, 50\% of the training data, removes the backdoor almost entirely, and a defender with that much data has already redone a large part of the outsourced training. All three defenses keep MAE$_\text{c}$ within 0.05 of the benign model, so clean forecasting performance cannot tell the victim whether the backdoor has been removed.
\begin{table}[t]
  \centering
  \caption{Backdoor removal rate (\%) under model-level defenses with TimesNet, averaged over three seeds.}
  \footnotesize
  \renewcommand{\arraystretch}{0.92} %
    \begin{tabular}{c|c|cccc}
    \toprule
    \multirow{2}{*}{Dataset} & \multirow{2}{*}{Defense} & \multicolumn{4}{c}{Clean data used by the defender} \\
    \cmidrule{3-6}
     & & 5\% & 10\% & 20\% & 50\% \\
    \midrule
    \multirow{3}{*}{ETTh1} & Pruning & 34.2 & 34.2 & 34.2 & 34.2 \\
     & Fine-tuning & 52.0 & 65.4 & 80.5 & 88.8 \\
     & Fine-pruning & 67.9 & 75.4 & 86.5 & 96.3 \\
    \midrule
    \multirow{3}{*}{ETTm1} & Pruning & 28.4 & 29.2 & 28.4 & 28.4 \\
     & Fine-tuning & 18.0 & 28.9 & 57.1 & 85.7 \\
     & Fine-pruning & 53.5 & 69.4 & 81.0 & 98.2 \\
    \midrule
    \multirow{3}{*}{Weather} & Pruning & 12.6 & 12.0 & 12.0 & 12.0 \\
     & Fine-tuning & 19.5 & 34.2 & 65.5 & 94.5 \\
     & Fine-pruning & 20.5 & 30.0 & 47.9 & 62.0 \\
    \bottomrule
    \end{tabular}%
  \label{tab:defense}%
\end{table}%

\subsection{Ablation Analysis}
In this subsection, we conduct ablation studies to evaluate the individual contribution of each key design in \name. We consider five ablation experiments for four data poisoning designs and the custom multi-objective loss in \name and evaluate them on the ETTh1 dataset and the TimesNet model, running each with the three random seeds 2025, 2026, and 2027. The variants are implemented as follows:
\begin{enumerate}
    \item \textbf{w/o-HPSS (Hybrid poisoned sample selection):} We first vary the similarity mixing ratio $\gamma$ from 0\% to 100\%, and then remove the hybrid sample selection strategy and randomly select poisoned samples.
    \item \textbf{w/o-GAT (GAT-based poisoned variable selection):} We replace the GAT-based poisoned variable selection with random selection.
    \item \textbf{w/o-Lag (Lag-aware trigger positioning):} We replace our lag-aware trigger insertion strategy with two alternatives: random insertion (w/o-Lag) and fixed insertion at the input sequence end (Input End).
    \item \textbf{w/o-Opt (Optimized trigger):} We disable the trigger optimization mechanism and use either fixed initialized triggers (InitOnly) or fixed random triggers (RandOnly). RandOnly follows the fixed-trigger design of BadNets~\cite{gu2017badnets} while keeping all other designs of \name.
    \item \textbf{w/o-CustomObj (Customized objective):} We replace our customized optimization objectives in Eqs.~\eqref{OptimizationM} and~\eqref{OptimizationT} with the standard backdoor training objective in Eq.~\eqref{ConO}.
\end{enumerate}
\begin{table}[t]
  \centering
  \vspace{0.0cm}
  \caption{Ablation results on ETTh1 with TimesNet (mean $\pm$ std over three seeds for \name and its variants). Benign and BackTime are single runs from Table~\ref{tab:comparison1}. In the top block, bold marks the stronger attack and the clean MAE closer to Benign; in the sample-selection block, it marks the lowest target MAE. Horizontal rules separate ablation modules.}
  \vspace{0.0cm}
  {\footnotesize\renewcommand{\arraystretch}{0.9} %
    \begin{tabular}{c|ccc}
    \toprule
     Method & MAE$_\text{c}\!{\downarrow}$  & MAE$_\text{pa}\!{\downarrow}$ & MAE$_\text{pn}\!{\downarrow}$ \\
    \midrule
    Benign & 0.424  & 0.781  & 0.432 \\
    BackTime~\cite{lin2024backtime} & 0.448 & 0.750 & 0.439 \\
    \rowcolor{RQFrame!12}\name & \textbf{0.437}$\pm$\textbf{0.003} & \textbf{0.312}$\pm$\textbf{0.026} & 0.445$\pm$0.005 \\
    \midrule
    $\gamma=0\%$ & 0.447$\pm$0.009 & 0.333$\pm$0.030 & 0.436$\pm$0.015 \\
    $\gamma=25\%$ & 0.438$\pm$0.004 & 0.307$\pm$0.027 & 0.425$\pm$0.004 \\
    $\gamma=50\%$ & 0.440$\pm$0.002 & \textbf{0.300}$\pm$\textbf{0.018} & 0.427$\pm$0.005 \\
    $\gamma=75\%$ & 0.443$\pm$0.003 & 0.317$\pm$0.019 & 0.433$\pm$0.005 \\
    $\gamma=100\%$ & 0.440$\pm$0.002 & 0.372$\pm$0.007 & 0.434$\pm$0.008 \\
    w/o-HPSS & 0.433$\pm$0.006 & 0.408$\pm$0.008 & 0.429$\pm$0.003 \\
    \midrule
    w/o-GAT & 0.490$\pm$0.056 & 0.589$\pm$0.092 & 0.480$\pm$0.049 \\
    \midrule
    w/o-Lag & 0.427$\pm$0.003 & 0.435$\pm$0.036 & 0.425$\pm$0.003 \\
    Input End & 0.434$\pm$0.004 & 0.445$\pm$0.032 & 0.426$\pm$0.004 \\
    \midrule
    InitOnly & 0.431$\pm$0.003 & 0.502$\pm$0.009 & 0.431$\pm$0.002 \\
    RandOnly & 0.504$\pm$0.007 & 0.525$\pm$0.018 & 0.454$\pm$0.011 \\
    \midrule
    w/o-CustomObj & 0.468$\pm$0.025 & 0.561$\pm$0.042 & 0.476$\pm$0.021 \\
    \bottomrule
    \end{tabular}%
    }
    \vspace{0.0cm}
  \label{tab:Ablation}%
\end{table}%
Table~\ref{tab:Ablation} reports the ablation results. Every variant is weaker than the full attack, and removing any single design raises MAE$_\text{pa}$ by at least 31\% and by up to 89\%. We group the five designs by the size of their contribution:
\begin{enumerate}
    \item \textbf{Core designs:} The GAT-based variable selection, the customized objective, and the trigger optimization determine whether the attack succeeds. Removing the GAT selection (\mbox{w/o-GAT}) spreads the trigger over variables with little influence on the target, which causes the largest degradation and the highest variance across seeds. Replacing our objective with the standard one (\mbox{w/o-CustomObj}) lets the clean samples dominate the gradient, so the model no longer binds the target pattern to the trigger. Freezing the trigger also weakens the attack, whether it is initialized from data (InitOnly) or at random (RandOnly), and since the data-driven initialization is stronger than a random one while optimization yields the largest gain, both the initialization and the optimization contribute. A random trigger additionally degrades clean forecasting.

    \item \textbf{Enhancement designs:} The hybrid sample selection and the lag-aware positioning refine how the backdoor is injected and give smaller gains on top of the core designs. Mixing label-similar and label-dissimilar windows is more effective than poisoning either kind alone, because a single kind lets the model attribute the target pattern to the shared features of that kind rather than to the trigger, consistent with Observation~1. Selecting windows at random (\mbox{w/o-HPSS}) is worse than any fixed mixing ratio, so controlling the ratio is what helps, and the attack stays effective across a broad range of ratios. Placing each trigger segment at the lag of its strongest correlation with the target outperforms both random placement (\mbox{w/o-Lag}) and the fixed input-end placement of prior work (Input End), which shows that when the trigger acts matters as much as which variables carry it.
\end{enumerate}

\begin{table}[t]
  \centering
  \footnotesize
  \renewcommand{\arraystretch}{0.9}
  \caption{Robustness of \name to the number of target variables and the target pattern shape. The notation $x/7$ denotes $x$ target variables out of 7 total variables.}
  \begin{tabular}{c|ccc}
    \toprule
    \multicolumn{4}{c}{(a) Number of target variables}\\
    \midrule
      & MAE$_\text{c}\!{\downarrow}$ & MAE$_\text{pa}\!{\downarrow}$ & MAE$_\text{pn}\!{\downarrow}$ \\
    \midrule
    1/7 (default) & 0.435 & 0.336 & 0.440 \\ %
    2/7 & 0.439 & 0.430 & 0.393 \\
    3/7 & 0.445 & 0.454 & 0.459 \\
    4/7 & 0.441 & 0.494 & 0.416 \\
    5/7 & 0.459 & 0.512 & 0.308 \\
    \midrule
    \multicolumn{4}{c}{(b) Target pattern shape}\\
    \midrule
    Hump (default) & 0.435 & 0.336 & 0.440 \\ %
    Upward trend & 0.425 & 0.375 & 0.429 \\
    Up and down & 0.417 & 0.378 & 0.422 \\
    \bottomrule
  \end{tabular}
  \label{tab:targetv}\label{tab:tpshape}
\end{table}
\subsection{Robustness Analysis}
\begin{figure}[t]
    \centering
    \includegraphics[width=\linewidth]{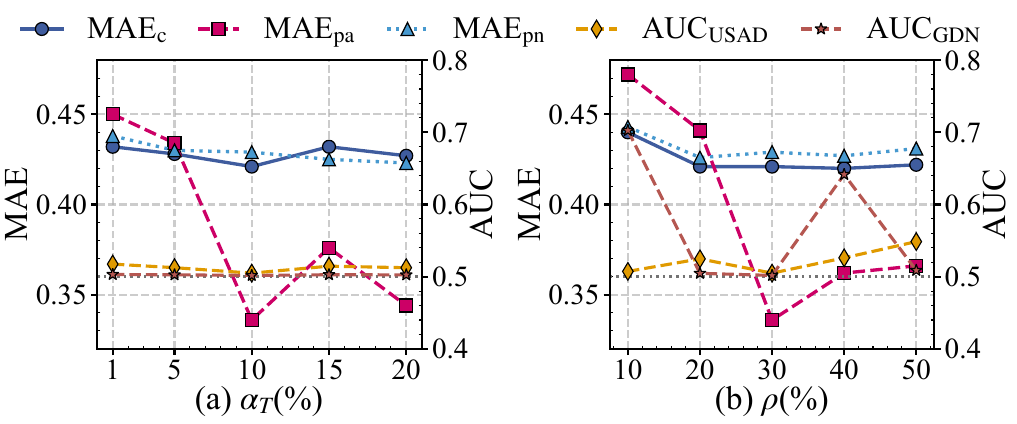}
    \caption{Attack performance of \name under varying (a) poisoning ratio $\alpha_T$ and (b) trigger length ratio $\rho$. MAE curves use the left y-axis, where lower is better, and AUC curves of USAD and GDN use the right y-axis, where the dotted line marks chance level.}
    \label{BadTime_metric_vs_alpha_both}
\end{figure}

\begin{figure}[t]
    \centering
    \includegraphics[width=\linewidth]{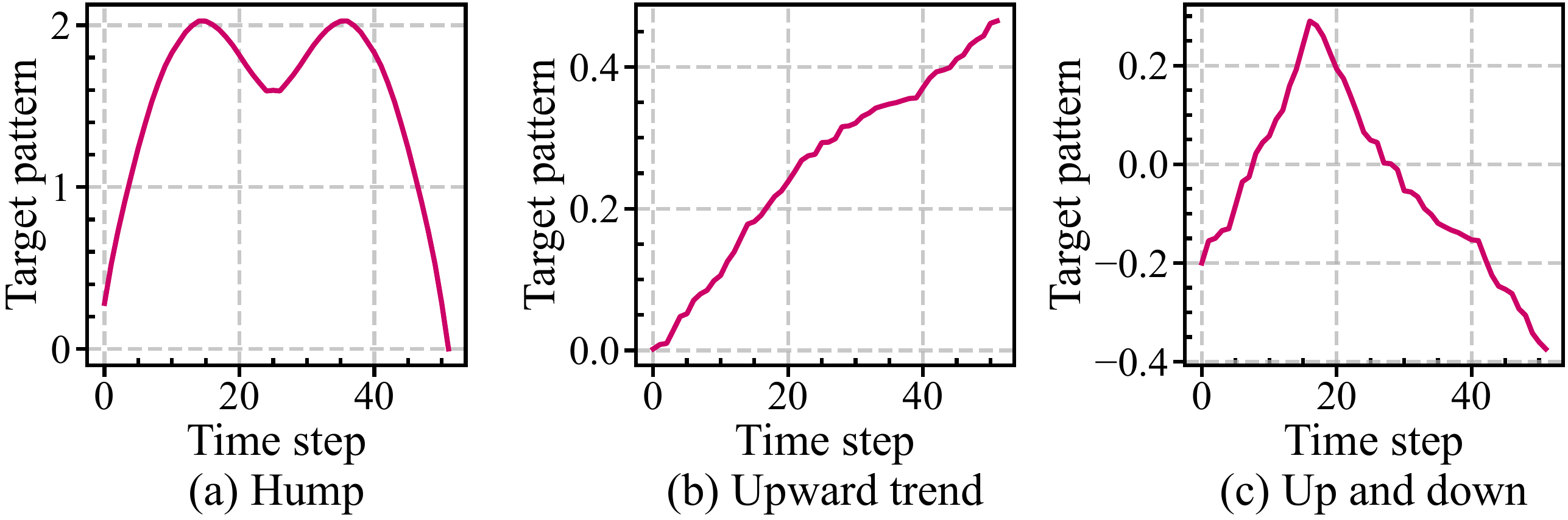}
    \caption{The three target pattern shapes evaluated in Table~\ref{tab:tpshape}.}
    \label{three_target_pattern}
\end{figure}

We assess the robustness of \name under varying experimental settings. In each case, only one parameter is modified while others follow the defaults in Table~\ref{symboltable2}. Fig.~\ref{BadTime_metric_vs_alpha_both}(a) shows that our method remains effective even when only 1\% of the training samples are poisoned. Increasing $\alpha_T$ introduces more poisoned samples, allowing the model to better learn the mapping between triggers and target patterns, thereby enhancing attack performance.  AUC$_\text{USAD}$ and AUC$_\text{GDN}$ stay around 0.5 at every $\alpha_T$, so a higher poisoning ratio does not make the trigger easier to detect. Fig.~\ref{BadTime_metric_vs_alpha_both}(b) shows that \name remains effective even when the trigger length is only 10\% of the input length.  Table~\ref{tab:targetv}(a) shows that the attack performance of \name gradually declines as the number of target variables increases, due to the growing difficulty of steering more variables toward predefined patterns, yet it remains effective even when targeting five of the seven variables. Table~\ref{tab:poisonedv} shows that the attack is strongest with two poisoned variables. A single poisoned variable gives a slightly weaker attack, consistent with the motivation of Design II to spread the trigger over correlated variables, and beyond two variables the effectiveness gradually declines because the fixed total trigger length is spread across more variables, yielding shorter, fragmented segments that are harder for the model to learn.  Table~\ref{tab:tpshape}(b) shows that \name maintains high attack effectiveness across all pattern types (visualized in Fig.~\ref{three_target_pattern}), demonstrating its strong robustness to variations in target pattern shape.
\begin{table}[t]
  \centering
  \fontsize{7pt}{7pt}\selectfont
  \caption{Performance of \name when attacking with different numbers of poisoned variables. The notation $x/7$ indicates $x$ poisoned variables out of 7 total variables. AUC is measured as in Table~\ref{tab:comparsion2}.}
  \renewcommand{\arraystretch}{0.9} %
  \resizebox{\linewidth}{!}{
    \begin{tabular}{c|ccccc}
    \toprule
          & MAE$_\text{c}\!{\downarrow}$  & MAE$_\text{pa}\!{\downarrow}$ & MAE$_\text{pn}\!{\downarrow}$ & AUC$_\text{USAD}$ & AUC$_\text{GDN}$ \\
    \midrule
     1/7  & 0.423  & 0.363  & 0.433  &  0.528  &  0.505  \\
     2/7 (default)  & 0.435  & 0.336  & 0.440  &  0.505  &  0.502  \\ %
     3/7  & 0.423  & 0.401  & 0.429  &  0.508  &  0.609  \\
     4/7  & 0.429  & 0.440  & 0.432  &  0.501  &  0.573  \\
     5/7  & 0.436  & 0.448  & 0.435  &  0.502  &  0.555  \\
    \bottomrule
    \end{tabular}%
    }
  \label{tab:poisonedv}%
\end{table}%

\begin{table}[H]
  \centering
  \caption{Per-epoch training cost with TimesNet on a single H100 GPU, where the trigger update is amortized over epochs.}
  \renewcommand{\arraystretch}{0.92} %
  \resizebox{\linewidth}{!}{
    \begin{tabular}{c|c|cccc|c}
    \toprule
    Dataset & Method & \makecell{Training\\(s)} & \makecell{Trigger\\update (s)} & \makecell{Total\\(s)} & Relative & \makecell{Memory\\(MB)} \\
    \midrule
    \multirow{3}{*}{ETTh1} & Benign & 6.06 & -- & 6.06 & 1.00$\times$ & 314 \\
     & BackTime~\cite{lin2024backtime} & 5.85 & 2.68 & 8.53 & 1.41$\times$ & 366 \\
     & \cellcolor{RQFrame!12}\name & \cellcolor{RQFrame!12}6.21 & \cellcolor{RQFrame!12}4.44 & \cellcolor{RQFrame!12}10.65 & \cellcolor{RQFrame!12}1.76$\times$ & \cellcolor{RQFrame!12}399 \\
    \midrule
    \multirow{3}{*}{Electricity} & Benign & 138.62 & -- & 138.62 & 1.00$\times$ & 3115 \\
     & BackTime~\cite{lin2024backtime} & 139.46 & 60.11 & 199.56 & 1.44$\times$ & 3831 \\
     & \cellcolor{RQFrame!12}\name & \cellcolor{RQFrame!12}216.46 & \cellcolor{RQFrame!12}174.14 & \cellcolor{RQFrame!12}390.60 & \cellcolor{RQFrame!12}2.82$\times$ & \cellcolor{RQFrame!12}4917 \\
    \bottomrule
    \end{tabular}%
    }
  \label{tab:overhead}%
\end{table}%

\subsection{Computational Overhead}
We measure the training cost of \name, BackTime, and clean training (Benign) with TimesNet on a single NVIDIA H100 GPU, using ETTh1 with 7 variables and Electricity with 321 variables. Table~\ref{tab:overhead} reports the per-epoch cost. The one-time preprocessing of Phase I takes only about 1.3~s on both datasets, and the main overhead comes from trigger optimization during training.  From ETTh1 to Electricity, the number of variables grows 46-fold, whereas the cost of \name relative to clean training grows only 1.6-fold, so the overhead scales well with the number of variables. Even on Electricity, one epoch takes less than seven minutes and uses less than 5 GB of GPU memory. The additional time of \name over BackTime is also mostly spent on trigger optimization, a core design that accounts for a large part of the attack strength in Table~\ref{tab:Ablation}. All of this cost is paid offline by the attacker who performs the outsourced training. Since the backdoored model keeps the architecture of the victim model, it incurs no additional cost at inference time.

\section{Conclusion}
\label{conclusion}

In this paper, we proposed \name, the first effective backdoor attack on MLTSF. Extending backdoor attacks to long horizons faces two challenges: a local trigger is attenuated over hundreds of forecast steps, and the backdoor signal is extremely sparse under a low poisoning rate. \name counters attenuation by distributing a data-initialized trigger over the variables that most influence the target, placing each segment at its estimated lag, and optimizing it alternately with the model. It counters sparsity with hybrid poisoned sample selection and a decoupled training objective that up-weights the backdoor signal. Experiments on seven datasets and six models show that \name reduces the MAE on the target variables by 57\% on average compared with BackTime while changing the average clean forecasting error by less than 1\%, extends the attackable horizon from 12 to 720 steps, and remains effective with only 1\% poisoned data. Removing any single design of \name raises the MAE on the target variables by at least 31\%, so every design contributes to this effectiveness.

\textbf{Limitation.} Pruning leaves the backdoor of \name largely intact, but fine-tuning on a large part of the training data can remove it. Backdoors that persist under such fine-tuning and defenses tailored to MLTSF are therefore important directions for future work. We hope this work draws attention to the security risks of outsourcing the training of MLTSF models deployed in critical applications.

\appendices

\section{Descriptions of Datasets}
\label{Datasets}
Table~\ref{tab:addlabel} summarizes the dimensions, complete-series lengths, and chronological train/validation/test splits of the seven datasets. For ETT, we follow TSLib and use the first 12/4/4 months, totaling 14,400 hourly observations or 57,600 observations at 15-minute intervals. The other datasets use a 7/1/2 split.

\textbf{ETTh1/ETTh2 and ETTm1/ETTm2 datasets~\cite{zhou2021informer}.} These Electricity Transformer Temperature datasets record six transformer load measurements and oil temperature from a Chinese power grid. Each ETTh series contains 17,420 hourly observations; each ETTm series contains 69,680 observations at 15-minute intervals.

\textbf{Electricity dataset~\cite{Autoformer}.} This dataset contains 26,304 hourly observations of electricity consumption for 321 customers from the UCI repository\footnote{https://archive.ics.uci.edu/dataset/321/electricityloaddiagrams20112014}.

\textbf{Weather dataset~\cite{Autoformer}.} This dataset contains 52,696 observations of 21 meteorological variables, sampled every 10 minutes throughout 2020.

\textbf{Exchange dataset~\cite{Autoformer}.} This dataset records 7,588 daily observations of the exchange rates of eight currencies against the U.S. dollar from 1990 to 2016.
\begin{table}[H] %
  \centering
  \fontsize{18pt}{18pt}\selectfont
  \vspace{0.0cm}
  \caption{Statistics and chronological splits of the seven datasets. Length denotes the complete series.}
  \vspace{0.0cm}
  \renewcommand{\arraystretch}{0.92} %
  \resizebox{\linewidth}{!}{
    \begin{tabular}{l|llll}
    \toprule
    Dataset & Dimension & Length & Domain & \makecell{Split \\(Train/Val/Test)} \\
    \midrule
    ETTh1, ETTh2 & 7     & 17,420 & Electricity (1 hour) & 6/2/2 \\
    ETTm1, ETTm2 & 7     & 69,680 & Electricity (15 min) & 6/2/2 \\
    Electricity & 321   & 26,304 & Electricity (1 hour) & 7/1/2 \\
    Weather & 21    & 52,696 & Environment (10 min) & 7/1/2 \\
    Exchange & 8     & 7,588 & Economic (1 day) & 7/1/2 \\
    \bottomrule
    \end{tabular}%
    }
  \label{tab:addlabel}%
\end{table}%

\section*{Acknowledgment}
This work was supported in part by the National Natural Science Foundation of China under Grant U24A20259; in part by the Key Laboratory of Computing Power Network and Information Security, Ministry of Education, Qilu University of Technology (Shandong Academy of Sciences) under Grant 2023ZD003; and in part by the Key Laboratory of Data Protection and Intelligent Management, Ministry of Education, Sichuan University under Grant SCUSAKFKT202403Y.

\bibliographystyle{IEEEtran}
\bibliography{reference.bib}

\end{document}